\documentclass[longauth,bibyear]{aa}
\usepackage{txfonts}
\usepackage{graphicx}   
\usepackage{amsmath}    
\usepackage{array} 
\usepackage{multicol,rotating} 
\usepackage{longtable} 
\usepackage[section]{placeins}
\usepackage{color}

\begin{document}

\newcommand{\abs}[1]{\ensuremath{\lvert#1\rvert}}

\title{Linking surface morphology, composition, and activity on the nucleus of 67P/Churyumov-Gerasimenko}
\subtitle{}

\author{S. Fornasier\inst{1}
\and V.H. Hoang\inst{1,2} 
\and P.H. Hasselmann\inst{1} 
\and C. Feller\inst{1} 
\and M.A. Barucci\inst{1} 
\and J.D.P. Deshapriya\inst{1} 
\and H. Sierks\inst{3} 
\and G. Naletto\inst{4, 5, 6} 
\and P. L. Lamy\inst{7} 
\and R. Rodrigo\inst{8,9} 
\and D. Koschny\inst{10} 
\and B. Davidsson\inst{11} 
\and J. Agarwal\inst{3} 
\and C. Barbieri\inst{4}
\and J.-L. Bertaux\inst{7} 
\and I. Bertini\inst{4} 
\and D. Bodewits\inst{12} 
\and G. Cremonese\inst{13} 
\and V.  Da Deppo\inst{6} 
\and S.  Debei\inst{14} 
\and M. De Cecco\inst{15} 
\and J. Deller\inst{3} 
\and S. Ferrari\inst{16} 
\and M. Fulle\inst{17} 
\and P. J. Gutierrez\inst{18} 
\and C. G\"uttler\inst{3} 
\and W.-H. Ip\inst{19,20} 
\and H.U. Keller\inst{21,22} 
\and  M. K\"uppers\inst{23} 
\and  F. La Forgia\inst{4} 
\and M. L. Lara\inst{18} 
\and M. Lazzarin\inst{4} 
\and Z-Y Lin\inst{19} 
\and J.J. Lopez Moreno\inst{18} 
\and F. Marzari\inst{4} 
\and S. Mottola\inst{21} 
\and M. Pajola\inst{4} 
\and X. Shi\inst{3} 
\and I. Toth\inst{24} 
\and C. Tubiana\inst{3}
}
%
\institute{LESIA, Observatoire de Paris, Universit\'e PSL, CNRS, Univ. Paris Diderot, Sorbonne Paris Cit\'{e}, Sorbonne Universit\'e, 5 Place J. Janssen, 92195 Meudon Pricipal Cedex, France \email{sonia.fornasier@obspm.fr}
\and Center for Technical Physics, Institute of Physics, Vietnam Academy of Science and Technology
\and Max-Planck-Institut f\"ur Sonnensystemforschung, Justus-von-Liebig-Weg, 3, 37077, G\"ottingen, Germany
\and University of Padova, Department of Physics and Astronomy {\it Galileo Galilei}, Via Marzolo 8, 35131 Padova, Italy 
\and Center of Studies and Activities for Space (CISAS) {\it G. Colombo}, University of Padova, Via Venezia 15, 35131 Padova, Italy
\and CNR-IFN UOS Padova LUXOR, Via Trasea, 7, 35131 Padova, Italy
\and Laboratoire Atmosph\`eres, Milieux et Observations Spatiales, CNRS \& Universit\'e de Versailles Saint-Quentin-en-Yvelines, 11 boulevard d'Alembert, 78280 Guyancourt, France
\and Centro de Astrobiologia, CSIC-INTA, 28850 Torrejon de Ardoz, Madrid, Spain
\and International Space Science Institute, Hallerstrasse 6, 3012 Bern, Switzerland
\and Scientific Support Office, European Space Research and Technology Centre/ESA, Keplerlaan 1, Postbus 299, 2201 AZ Noordwijk ZH, The Netherlands
\and Jet Propulsion Laboratory, M/S 183-401, 4800 Oak Grove Drive, Pasadena, CA 91109, USA
\and Department of Astronomy, University of Maryland, College Park, MD 20742-2421, USA
\and INAF, Osservatorio Astronomico di Padova, Vicolo dell'Osservatorio 5, 35122 Padova, Italy
\and University of Padova, Department of Mechanical Engineering, via Venezia 1, 35131 Padova, Italy
\and University of Trento, Faculty of Engineering, Via Mesiano 77, 38100 Trento, Italy
\and Dipartimento di Geoscienze, University of Padova, via G. Gradenigo 6, 35131 Padova, Italy
\and INAF Astronomical Observatory of Trieste, Via Tiepolo 11, 34014 Trieste, Italy
\and Instituto  de Astrof\'isica de Andalucia (CSIC), c/ Glorieta de la Astronomia s/n, 18008 Granada, Spain
\and National Central University, Graduate Institute of Astronomy, 300 Chung-Da Rd, Chung-Li 32054, Taiwan
\and Space Science Institute, Macau University of Science and Technology, Avenida Wai Long, Taipa, Macau 
\and Deutsches Zentrum f\"ur Luft und Raumfahrt (DLR), Institut f\"ur Planetenforschung, Asteroiden und Kometen, Rutherfordstrasse 2, 12489 Berlin, Germany
\and Institut f\"ur Geophysik und extraterrestrische Physik (IGEP), Technische Universitat Braunschweig, Mendelssohnstr. 3, 38106 Braunschweig, Germany
\and Operations Department, European Space Astronomy Centre/ESA, P.O.Box 78, 28691 Villanueva de la Canada, Madrid, Spain
\and MTA CSFK Konkoly Observatory, Budapest, Hungary 
}

\date{Accepted 27 August 2018. Received on July 2018}

\newpage

 \abstract{}{The Rosetta space probe accompanied comet 67P/Churyumov-Gerasimenko for more than two years, obtaining an unprecedented amount of unique data of the comet nucleus and inner coma. This has enabled us to study its activity almost continuously from 4 au inbound to 3.6 au outbound, including the perihelion passage at 1.24 au. This work focuses identifying the source regions of faint jets and outbursts and on studying the spectrophotometric properties of some outbursts. We use observations acquired with the OSIRIS/NAC camera during July-October 2015, that is, close to perihelion.}
{We analyzed more than 2000 images from NAC color sequences acquired with 7-11 filters covering the 250-1000 nm wavelength range. The OSIRIS images were processed with the OSIRIS standard pipeline up to level 3, that is, converted in radiance factor, then corrected for the illumination conditions. For each color sequence, color cubes were produced by stacking registered and illumination-corrected images.}
{ More than 200 jets of different intensities were identified directly on the nucleus. Some of the more intense outbursts appear spectrally bluer than the comet dark terrain in the vivible-to-near-infrared region. We attribute this spectral behavior to icy grains mixed with the ejected dust.\\  Some of the jets have an extremely short lifetime. They appear on the cometary surface during the color sequence observations, and vanish in less than some few minutes after reaching their peak. 
We also report a resolved dust plume observed in May 2016 at a resolution of 55 cm/pixel, which allowed  us to estimate an optical depth of $\sim$0.65 and an ejected mass of $\sim$ 2200 kg, assuming a grain bulk density of $\sim$ 800 kg/m$^3$.

We present the results on the location, duration, and colors of active sources on the nucleus of 67P from the medium-resolution (i.e., 6-10 m/pixel) images acquired close to perihelion passage. The observed jets are mainly located close to boundaries between different morphological regions. Some of these active areas were observed and investigated at higher resolution (up to a few decimeter per pixel) during the last months of operations of the Rosetta mission.} {These observations allow us to investigate the link between morphology, composition, and activity of cometary nuclei. Jets depart not only from cliffs, but also from smooth and dust-covered areas, from fractures, pits, or cavities that cast shadows and favor the recondensation of volatiles. This study shows that faint jets or outbursts continuously contribute to the cometary activity close to perihelion passage, and that these events are triggered by illumination conditions. Faint jets or outbursts are not associated with a particular terrain type or morphology. 
}

\titlerunning{Jets and activity on comet 67P }
   \maketitle
\section{Introduction}

\begin{table*}
         \begin{center} 
         \caption{Observing conditions for the NAC images ($\alpha$ is the  phase angle, r$_h$ is the heliocentric distance, and $\Delta$ is the distance between comet and spacecraft). Each sequence consists of images acquired with the 11 filters of the NAC camera:  F22 (649.2 nm), F23 (535.7 nm), F24 (480.7 nm), F16 (360.0 nm), F27 (701.2 nm), F28 (743.7 nm), F41 (882.1 nm), F51 (805.3 nm), F61 (931.9 nm), F71 (989.3 nm), and F15 (269.3 nm). 
$^1$: Number of analyzed sequences on the date. The sequences usually cover slightly more than one rotation period of the comet.
$^2$: This entry specifically refers to the number of jets that have been successfully located on the surface, not to the total number of jets. Some jets originate from behind the limb, and we cannot precisely locate their source region.
$^3$: This is the average number of detected jets per sequence for a given date.}
         \label{obs}
        \begin{tabular}{|c|c|c|c|c|c|c|c|} \hline
\bf{2015 } & \boldmath{N$_{sequences}^{1}$} & \boldmath{N$_{jets}^{2}$} & \boldmath{$\alpha  ~(^{\circ})$}  & \boldmath{r$_{h}$} {\bf(AU)} & \boldmath{$\Delta$} {\bf(km)} & \bf{Res. (m/px)} & \bf{Avg.}\boldmath{$^3$} \\ \hline
June 27   & 24  &19 &89.3--89.7   &1.365 &182.8--198.3 & 3.4--3.7 &0.8\\
July 26   & 33  &56 &89.9          & 1.262 &167.3--169.2 &3.2 &1.7\\
August 1  & 29  &101& 89.4--89.7  & 1.251& 206.4--215.& 4.0& 3.4\\
August 9  & 31  &19 &89.0--89.2   &1.244  &303.9--310.0 &5.8 &0.5\\
August 12 & 23  &10 &89.4--89.7   &1.243 &327.2--336.2 &6.3--7.0 &0.4\\
August 23 & 16  &31 &87.3--88.5   &1.250 &329.9--336.7 &6.2--6.3 &1.9 \\
August 30 & 24  &126&70.0--70.2   &1.261 &402.5--405.1 &7.6 &5.3\\
September 5 & 15 &26 & 99.7--102.6& 1.276& 393.3--441.5& 7.4--8.3 &1.7\\
October 11& 12  &16 &60.9--61.5   &1.437 &520.1--529.2& 9.7--9.9 &1.3\\
October 21& 12  & 26&64.0--64.4   &1.487 &420.1--422.5& 7.9 &2.2\\
October 31& 13  &42 &61.6--63.0   &1.565 &287.2--305.0& 5.4--5.7& 3.2\\ \hline 
       \end{tabular}
\end{center}
 \end{table*}
The Rosetta mission of the European Space Agency was launched on 2 March 2004 to perform the most detailed study ever attempted of a comet. After ten years of interplanetary cruising, Rosetta entered the orbit of its primary target, the short-period comet 67P/Churyumov-Gerasimenko (hereafter 67P), in August 2014 and followed the comet for more than two years until 30 September 2016, when it landed on the surface of the nucleus.\\
Rosetta carried a broad suite of instruments, including the Optical, Spectroscopic, and Infrared Remote Imaging System (OSIRIS), which acquired more than 75000 images of the comet during the mission. OSIRIS is composed of two cameras: the Narrow Angle Camera (NAC) for nucleus surface and dust studies, and the Wide Angle Camera (WAC) for the wide-field coma investigations (see Keller et al., 2007, for further details).
OSIRIS enabled extensive studies at high resolution (down to 10 cm/pixel, and even better during the  final phase of Rosetta's descent) of the nucleus with several filters in the 250-1000 nm range.  OSIRIS also provided high-resolution images of the cometary activity and its evolution from 4 au inbound to 3.6 au outbound.  \\
The nucleus of 67P is bilobed, has a low density of 537.8$\pm$0.7 kg/m$^{3}$, and a high porosity (70-80\%, Sierks et al., 2015; P\"atzold et al., 2016; Jorda et al., 2016; Preusker et al., 2017). The surface is dark, with a geometric albedo of 5.9\% at 535 nm (Fornasier et al. 2015), and it shows a complex morphology that is characterized by consolidated and smooth terrains, depressions, pits, extensive layering, ubiquitous boulders, and dust-covered areas (Thomas et al., 2015, El-Maarry et al., 2015, 2016, Massironi et al., 2015). 
 
The coma activity of 67P has been monitored by several instruments even before the Rosetta rendezvous maneuver with the comet in August 2014. The OSIRIS images captured an outburst at the end of April 2014, when the comet was at 4 au and was not yet resolved by the cameras (Tubiana et al., 2015a), followed shortly after by the detection of water vapor with the MIRO instrument (Gulkis et al., 2015). During the first resolved observations, most of the activity arose from Hapi, the northern region located between the two lobes of the comet, which is brighter than average and relatively blue (Fornasier et al., 2015), and water ice and the first evidence of a diurnal water cycle was reported (de Sanctis et al., 2015). Important diurnal and seasonal variations were observed in the coma for different outgassing species. These were related to the complex morphology and the illumination conditions (Bockelee-Morvan et al., 2015, 2016, 2017; Biver et al., 2015;  Hansen et al., 2016; Luspay-Kuti et al., 2015; Lin et al., 2015, 2016; Lara et al., 2015). 

The OSIRIS instrument observed different activity events during the two years of continuous observations of the comet, allowing scientists to retrieve the positions on the surface of the nucleus of several jets through geometrical tracing, to photometrically characterize them, and to study their seasonal evolution  (Vincent et al., 2016a, 2016b; Lara et al., 2015 ; Lin et al. 2015, 2016, 2017; Shi et al., 2016, 2018). In particular, several peculiar events were investigated:  Shi et al. (2016) analyzed a cluster of sunset jets from the Ma'at region in late April 2015; Knollenberg et al. (2016) studied an outburst originating from a part of the Imhotep region on 12 March 2015; Vincent et al. (2016a) located and classified 34 outbursts that occurred between July and September 2015; Vincent et al. (2016b) observed that most outbursts were located near collapsed cliffs, which they interpreted as evidence of mass wasting; Pajola et al. (2017) reported the first unambiguous link between an outburst and a cliff collapse that they observed in the Aswan site, which is located in the Seth region, with direct exposure of the fresh icy interior of the comet; and Agarwal et al. (2017) reported an outburst event in the Imhotep region at 3.32 au outbound that altered an area with a radius of 10 m on the surface and left an icy patch. 

Most of the results on cometary activity studies are obtained from observing sequences with long exposure times that are devoted to investigating the faint cometary gas and dust emissions. In such observations, the nucleus is usually saturated. In these studies, the jet sources cannot be directly identified on the nucleus, and their location is retrieved by triangulation from different viewing geometries or from projecting the 2D jet coordinates on synthetic images of the nucleus at the time of a given observation. Moreover, Shi et al. (2018) investigated the relation between jet morphology and terrain and cautioned that the trace-back analysis of jets may be hindered by the observing geometry.

This  work focuses on the jets observed during the OSIRIS/NAC color sequences that are dedicated to studying the colors and composition of the nucleus. They were taken between June and October 2015, when the comet was immediately before and after perihelion, which occurred on 13 August 2015. Several jets were observed on the nucleus, which allows us to precisely locate them directly on the surface and to investigate the morphology of the source regions from images with higher resolution that were acquired later. These events include the perihelion outburst, several faint jets, and short-lived transient events that lasted for a few minutes.  \\
In section 2 we summarize the observational sets and the analysis we performed to characterize and localize the jets, and in section 3 we present the results of our jet distribution analysis over the southern hemisphere of the nucleus and describe their main properties. In section 4 we discuss the morphology of the jet sources from images with higher resolution that were acquired in 2016. Finally, we discuss the main mechanisms that are at the origin of activity and examine the link between morphology, composition, and activity of 67P.

\section{Observations and data analysis} 

\begin{figure*}
\centering
\includegraphics[width=1.0\textwidth]{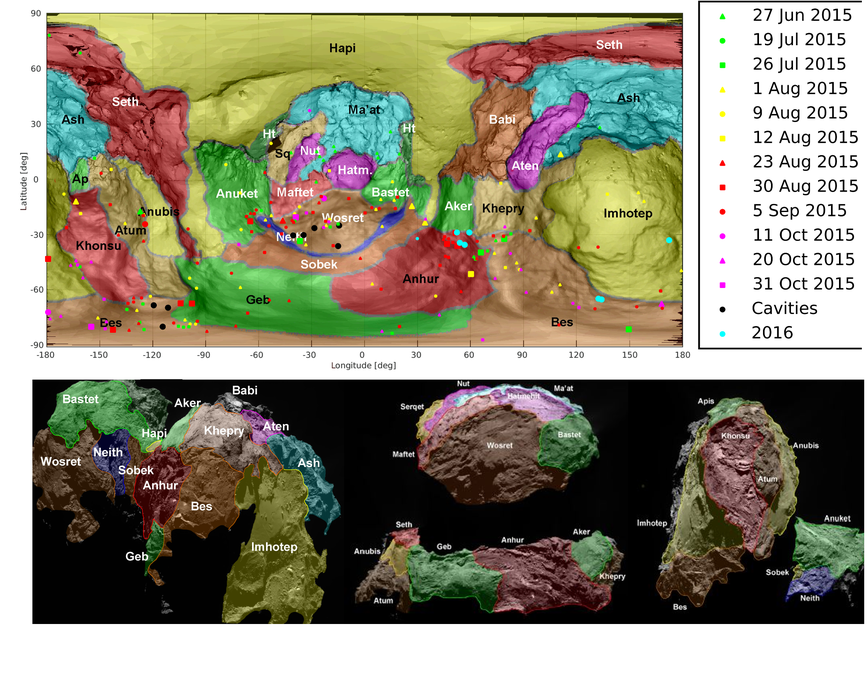}
\caption{Top: Map of the comet. We show superposed the jet sources that were identified during summer in the southern hemisphere of the comet. The locations of some nearby jets are averaged for clarity, and some notable jets are represented with larger symbols. Black circles represent cavities that were found to be active in different data sets (see Table~\ref{all_jets} for details). Cyan points represent events observed in 2016 that are reported here for the dust plume in the Bes region, are reported in Agarwal et al. (2017) for the outburst in the Imhotep region, and are reported in Fornasier et al. (2017) for the jets in the Anhur region. Bottom: Three different views of the  southern hemisphere of 67P. Regional boundaries are overlaid. The complete 3D views of the comet nucleus with all the regions superposed are shown in El-Maarry et al. (2015, 2016).}
\label{map}
\end{figure*}
\begin{figure*}
\centering
\includegraphics[width=0.95\textwidth,angle=0]{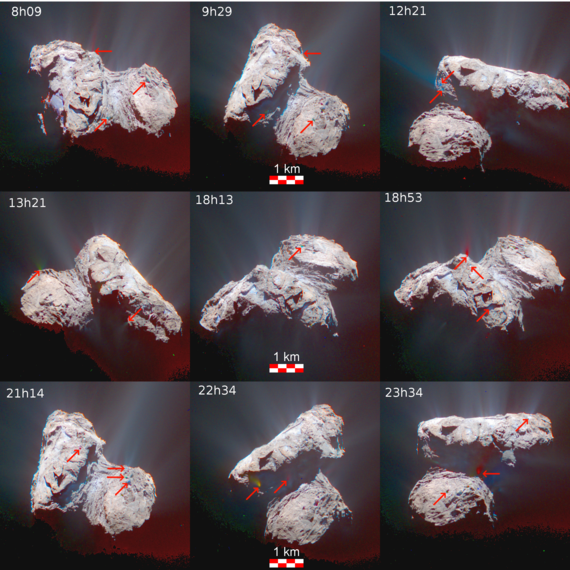}
\caption{RGB composite (from the images acquired with filters centered at 882, 649, and 480 nm) of 9 out of the 23 sequences acquired on 30-31 August 2015, two weeks after perihelion passage. Several faint jets are visible in different images, as well as local bright patches that are associated with the exposure of volatiles.}
\label{30aug2015}
\end{figure*}

We used color sequences devoted to the nucleus spectrophotometric characterization acquired with several filters of the NAC camera close to perihelion passage, between June and October 2015 (Table~\ref{obs}). Even if not expressly devoted to coma and dust studies, these sequences, acquired at the peak of cometary activity, revealed several jets and  outbursts. We thus decided to build a jet catalog based on these observations. The advantage of these sequences is that in contrast to the long-exposure time observations that are devoted to investigating the cometary activity, the nucleus is not saturated. This permits precisely locating the position of the activity sources on the nucleus and studying the colors of some outburst. In particular, these observations for the first time highlight short-lived jets with durations shorter than a few minutes. \\
 We analyzed more than 2000 images that were obtained over 11 days between June and October 2015. Each observing data set at a given date consists of 12-33 individual sequences of 11 filters (Table~\ref{obs}) and covers the rotational period of the nucleus (12.4 hours, Mottola et al., 2014). The observed nucleus surface is situated mostly in the southern hemisphere and on the comet equator. The identified jets therefore only account for a fraction of the total number of jets departing from the whole nucleus. \\
We used level 3B images from  the OSIRIS pipeline, which are corrected for bias, flat field, geometric distortion, calibrated in absolute flux (in $W~m^{-2}~ nm^{-1}~ sr^{-1}$), and are finally converted into radiance factor (called $I/F$, where I is the observed scattered radiance and F is the incoming solar irradiance at the heliocentric distance of the comet, divided by $\pi$), as described in Tubiana et al. (2015b) and Fornasier et al. (2015). \\
All images of an individual sequence were first coregistered using the F22 NAC filter (centered at 649.2 nm) for reference. For the coregistration, we used a python script based on the scikit-image library (Van der Walt et al., 2014), and the optical flow algorithm (Farneb\"ack, 2003), as has been done previously for the analyses presented in Fornasier et al. (2017) and Hasselmann et al. (2017). \\
Each image was reconstructed for illumination and observing geometry using the 3D stereo-photoclinometric shape model (Jorda et al., 2016), considering all relevant geometric parameters, such as the camera distortion model, the alignment of the instrument to the Rosetta spacecraft, and the orientation of the spacecraft (with reconstructed orbit position and pointing) with respect to the 67P nucleus and to the Sun.

RGB (Red, Green, and Blue) in false-color maps were generated from coregistered NAC images that were acquired with the filters centered at 882 nm, 649 nm, and 480 nm using the STIFF code (Bertin, 2012).
These RGB maps offer the first visual clues about the comet nucleus. In these images, most of the comet nucleus appears gray, and bright spots are displayed as white patches. Transient events, on the other hand, are usually displayed as colored areas as they are only captured by some of the filters during a sequence, or because their intensity peaks in some filters. \\
Finally, for the spectral analysis,  images were photometrically corrected by applying the Lommel-Seeliger disk law ($D$), which has been proven to correct satisfactorily for dark surfaces (Li et al., 2015):  
\begin{equation}
D(i,e) = \frac{2\mu_{i}}{\mu_{e}+\mu_{i}}
,\end{equation}
where $\mu_{i}$ and $\mu_{e}$ are the cosine of the solar 
incidence (i) and the emission (e) angles, respectively. The reflectance (at the phase angle of a given observation) of selected regions of interest (ROI) was computed from the photometrically corrected images by integrating the signal in a box of 3$\times$3 pixels, and the relative reflectance obtained by normalizing the spectrophotometry in the green filter, centered at 535 nm. The spectral slopes were evaluated in the 535-882 nm range, as detailed in Fornasier et al. (2015, 2016).

The jets were first identified in the RGB images as colored patches. The Cartesian coordinates (x, y, z) of their sources on the nucleus were obtained through images that were simulated from the shape model, and were converted into longitudes and latitudes as follows: 
\begin{equation}
lon(x,y) = arctan2(y,x) ~ ,~ \hspace{1cm} ~ lat = arctan{\frac{z}{\sqrt{x^2+y^2}}}
.\end{equation}
We used the Cheops reference frame described in Preusker et al. (2015) to retrieve the coordinates of the jet footprints. The reference of this frame is a boulder called Cheops in the Imhotep region, whose location is defined to be at longitude 142.35$^{\circ}$, latitude -0.28$^{\circ}$, and at a radial distance of 1395 m from the center of the nucleus.

%
                          \section{Location and properties of jets and outbursts}


More than 200 activity events have been identified in June-October 2015 from multi-filter images that were devoted to characterizing the nucleus (Table~\ref{all_jets}). We stress that this list is incomplete. Some jets originate from behind the limb, such that we cannot precisely locate their source region, and they are not considered in this study. Moreover, several jets were reported by other studies from long-exposure time sequences that were expressly devoted to investigating the activity; they are not considered in the following analysis either (Vincent et al., 2016a; Knollenberg et al., 2016; Shi et al., 2016; Schmitt et al., 2017; Lin et al., 2016, 2017). Table~\ref{all_jets} reports all the jet locations we identified in the nucleus color sequences that we analyze here, together with their type, repeatability, cometary local time, and a short description. The jet types are given on the basis of their shape, following the classification of Vincent et al. (2016a): A is a collimated jet, B is a wide plume, and C is a complex shape (broad and collimated). \\
\begin{figure*}
\centering
\includegraphics[width=0.95\textwidth,angle=0]{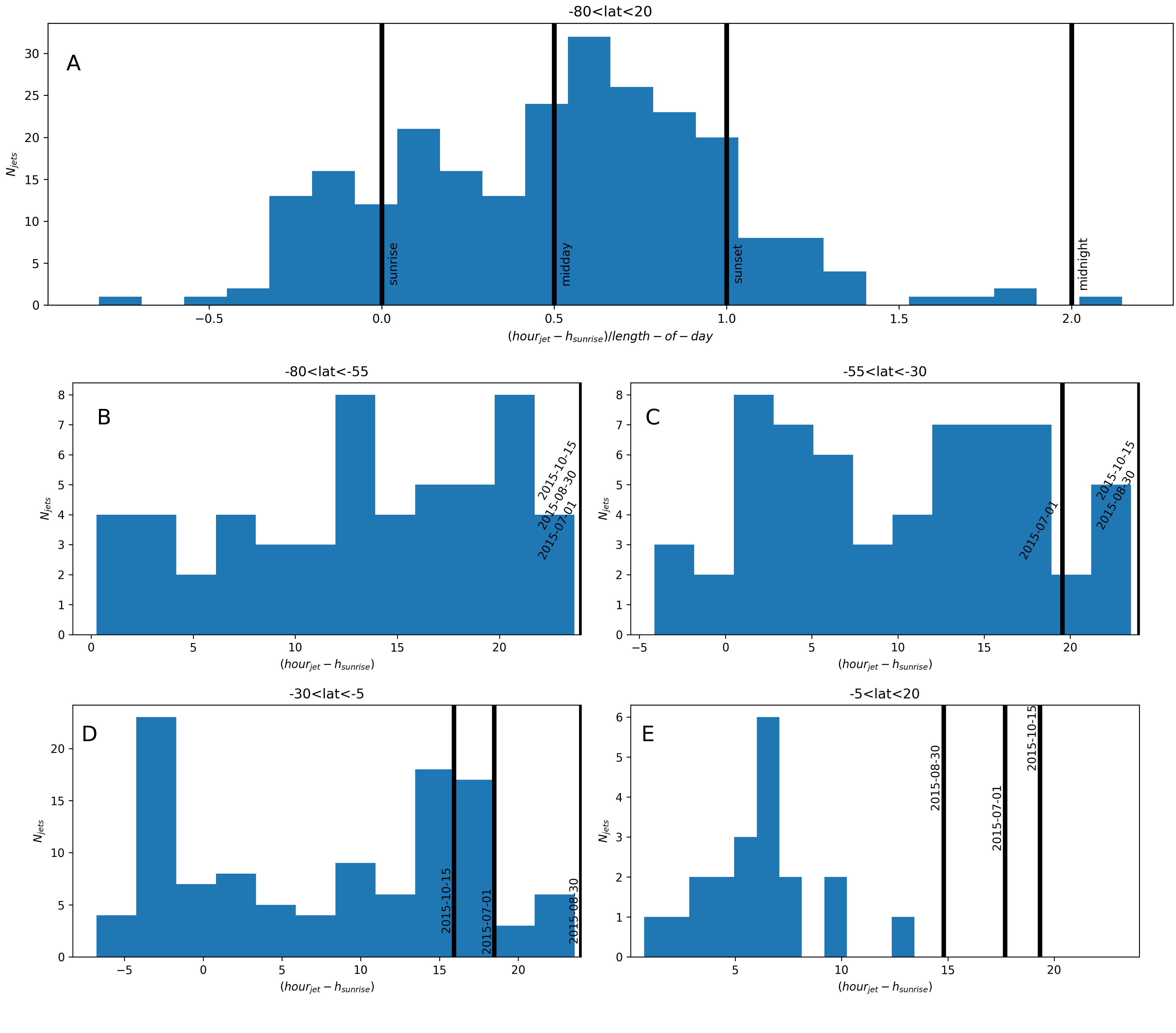}
\caption{A: Total number of observed jets observed as a function of the time since sunrise normalized per local day-time duration. Panels B-E: number of jets for four different latitude ranges as a function of the time since sunrise. The vertical black lines represent the approximate sunset time computed for three different dates near perihelion passage. The high southern latitudes in panel B were always illuminated in this period, and in this case, the sunrise time was set to zero. }
\label{local_time}
\end{figure*}
The jet positions are represented in Fig.~\ref{map} in a map of the nucleus showing the different morphological regions (El-Maarry et al., 2015, 2016). Most of the jets are close to the boundaries that separate the different morphological regions, where textural and topographic discontinuities are observed. These boundaries are between Sobek and Hapi, Sobek and Anuket, Wosret and Maftet, Wosret and Bastet, Anhur and Bes, Khepry and Bes, Anhur and Aker, Bes and Geb, Bes and Atum, and Bes and Khonsu. The association between jet location and the boundaries of the morphological regions has been reported by Vincent et al. (2016a), who found a clustering of the activity in the boundaries between Anhur and Aker and Anuket and Sobek on the large lobe, and in the boundary between Wosret and Maftet on the small lobe. \\
Some examples of jets observed on 30 August 2015 are reported in Fig.~\ref{30aug2015}. These images were acquired from relatively large distances, and thus the spatial resolution was low (about 7.6 m/px). However, the spatial extension of the jet sources was several pixels, which means that it is similar to or larger than the dust plume that was observed closely by the Rosetta instruments on 3 July 2016 from a distance of 8.5 km (see Fig. 2 in Agarwal et al., 2017).
\begin{figure*}
\centering
\includegraphics[width=0.85\textwidth,angle=0]{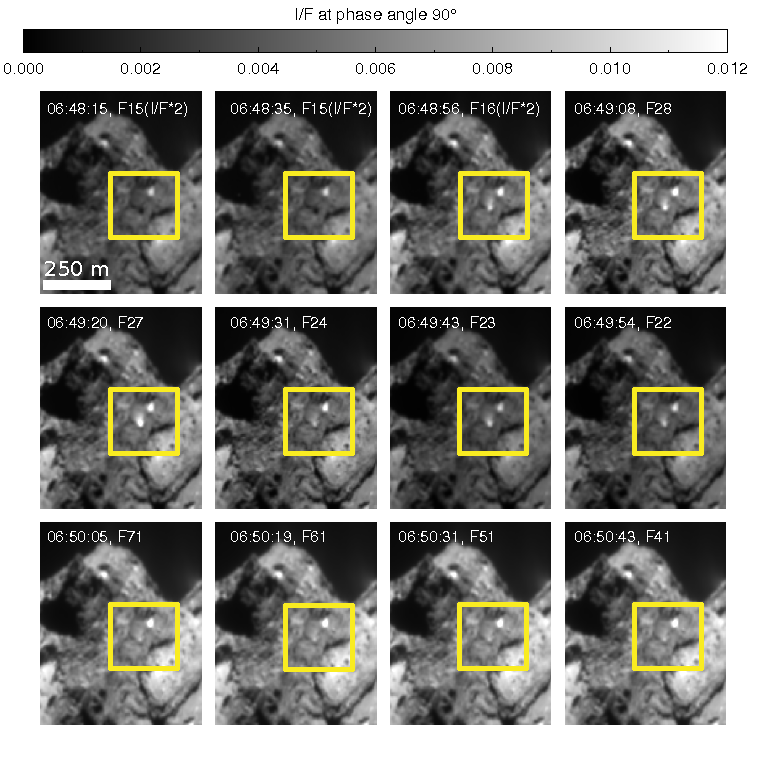}
\caption{Short-lived jet identified in OSIRIS NAC images at 6h48-6h50 on 30 August 2015 (jet 71 in Table~\ref{all_jets}). This sequence captures the beginning, peak, and end of the transient event, which lasted for approximately 90 seconds.}
\label{jet30aug}
\end{figure*}

\begin{figure*}
\centering
\includegraphics[width=0.95\textwidth,angle=0]{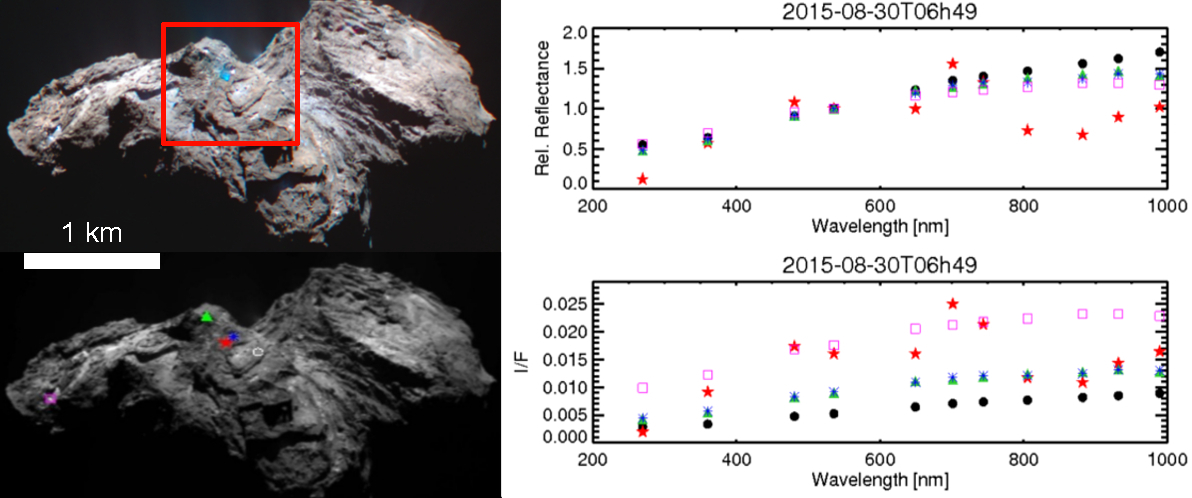}
\caption{Analysis of the short-lived jet identified in OSIRIS NAC images at 6h48-6h50 on 30 August 2015 (jet 71 in Table~\ref{all_jets}). Top left: RGB image (composed from filters centered at 882, 649, and 480 nm). The red square indicates the zoom into the area presented in Fig.~\ref{jet30aug}. Bottom left: Image acquired with the F22 filter. Five selected ROIs are superposed (the black circle shows the dark terrain, the red star represents the jet, and the blue asterisk, green triangle, and magenta square indicate different bright patches. Right plot: I/F factor (given at phase angle = 70$^{\circ}$) and relative reflectance (normalized at 535 nm) of the five selected ROIs.}
\label{jet30aug_anal}
\end{figure*}

 Several jets periodically originated from the same location inside cavities or alcoves (black circles in Fig.~\ref{map}), especially in Wosret and Bes. The walls of these cavities cast shadows that allowed the recondensation of volatiles. Evidence of exposed water ice has indeed been found inside them (see, e.g., Figs.~\ref{jet30aug}, ~\ref{jet30aug_anal}, and ~\ref{jet30aug_18h53}). The latitude and longitude position of jets departing from cavities is listed in Table~\ref{all_jets}. The errors indicate the range in longitude and latitude associated with periodic or close-by jets departing from these regions. These structures were active in several sequences, up to 61 times for cavity {\it A} in Wosret, and 29-47 times for cavities {\it A} and {\it B} in Bes. \\
While the perihelion sequence has the fewest observed jets per sequence (Table~\ref{obs}), it has the most spectacular and brightest event (jet 8 in Table~\ref{all_jets}). It originates from the Anhur region (Fornasier et al., 2017), and its intensity surpasses that of all other jets (Vincent et al., 2016a). 

       The activity peak, defined as the highest number of jets per sequence, from the observations we investigate here occurs on 30 August 2015 (Table~\ref{obs}). This agrees with results on the entire cometary activity as observed from the ground and from other ROSETTA instruments, which reported an activity peak in 67P  approximately two weeks after perihelion (Snodgrass et al., 2016). The highest water production rate  as found by ROSINA occurred 18-22 days after perihelion (Hansen et al., 2016).  Bockelee-Morvan et al. (2016) reported an abrupt increase in the water production of 67P six days after perihelion from coma observations with the VIRTIS instrument. The coma observations immediately after perihelion with VIRTIS also revealed an increase by a factor 2 for the CO$_2$, CH$_4$, and OCS abundances relative to water. Bockelee-Morvan et al. (2016) attributed this activity post-perihelion to the sublimation of volatile-rich layers near the surface. The exposure of volatile-rich layers close to perihelion was also reported by Fornasier et al. (2016) from OSIRIS observations of the nucleus colors and spectrophotometry.  \\
  The post-perihelion activity peak is due to thermal lag and to the low thermal inertia of the nucleus surface layers (10-30 or 10-50 J~K$^{-1}$~m$^{-2}$~s$^{-0.5}$, according to Schloerb et al. (2015) and Gulkis et al., (2015), respectively). VIRTIS spectrometer data showed that dust-layered areas have low thermal inertia (I), while the rougher consolidated terrain revealed higher thermal inertia,
I $\ge$ 50 J~K$^{-1}$~m$^{-2}$~s$^{-0.5}$ (Leyrat et al., 2015), such as the Abydos landing site, whose thermal inertia is 85$\pm$35 J~K$^{-1}$~m$^{-2}$~s$^{-0.5}$, as determined by in situ measurements with the MUPUS instrument on board the Philae lander (Spohn et al., 2015). 
These results suggest that the nucleus has a low thermal conductivity and that it is a highly porous body with a subsurface layer of dust and ice that locally has a highly compressive strength (Spohn et al., 2015).

Figure~\ref{local_time} shows the distribution of all the observed jets as a function of the comet local time evaluated on a 24-hours basis. Local time has been computed using the Rosetta NAIF-SPICE ESA kernels (Acton et al., 2016), which include all the geometrical information about the spacecraft and the positions of the comet and the Sun, assuming a rotational period of the comet of 12.4047 hours (Mottola et al., 2014). The illumination is computed from the local time of a given position, disregarding topography, that is to say, we did not consider mutual shadowing.
At the top of Figure~\ref{local_time} (panel  A), we present the distribution of all jets per local time from sunrise, normalized by the length of day (i.e., the time from sunrise to sunset). As the jets cover different latitudes over four months, the length of day strongly depends on latitude and epoch. The length of day was computed for each source, counted once in Figure~\ref{local_time}, for a given time. This normalization was needed to present the active sources in the same cometary time-frame with respect to their illumination time. In this way, sunrise corresponds to zero and sunset to one in panel A of Fig.~\ref{local_time}. In this plot, the majority of the jet sources is active during cometary afternoon, and few events take place around midnight. This behavior may be explained by the thermal lag needed to penetrate the layer of subsurface volatiles and activate sublimation. 
\\
We also report in Fig.~\ref{local_time} (panels B-E) the distribution of jet sources per different latitude range as a function of time since sunrise. In these plots we did not normalize by the length of day, but we indicate the approximate sunset time for three different dates.   
For regions close to the south pole (panel B in Fig.~\ref{local_time}), which are always illuminated during the time-frame we considered (we set the sunrise time to zero in this case), the majority of the sources is active after midday. The  medium-to-high southern latitudes (panel C), which mostly correspond to active sources in the Bes, Anhur, and Khepry regions, show a bimodal distribution with two activity peaks, one in the morning and one in the afternoon. The equatorial southern sources (panel C), which are mostly located in the Wosret region in the small lobe, also display a bimodal distribution, but with different peaks: one at night, a few hours before dawn, and one during sunset. Conversely, equatorial northern sources (panel E) display most of the activity about 5-7 hours after sunrise and show no events at sunset or during the night. \\
 The activity peaks in the afternoon, close to sunset, or during the night may be explained by the thermal lag to activate the sublimation of subsurface volatiles. Sunset jets have previously been observed, for instance, in the Ma'at region (Shi et al., 2016) or at night (Knollenberg et al., 2016). Conversely, more than half of the 34 outbursts observed by Vincent et al. (2016a) during the 67P summer occurred at dawn or early morning. This was interpreted as a consequence of rapid temperature variations that cause the surface to crack. For the fainter jets we report here, only sources at medium-to-high southern latitudes (panel C in Fig.~\ref{local_time}) present a maximum close to sunrise and early morning. The sublimation of recondensing frost or ice during the short summer night, which is periodically visible on the surface close to perihelion passage (Fornasier et al., 2016), may be an alternative explanation for some of the morning jets that were observed in the Bes-Anhur regions at these latitudes.

\subsection{Short-lived jets}

\begin{figure*}
\centering
\includegraphics[width=0.95\textwidth,angle=0]{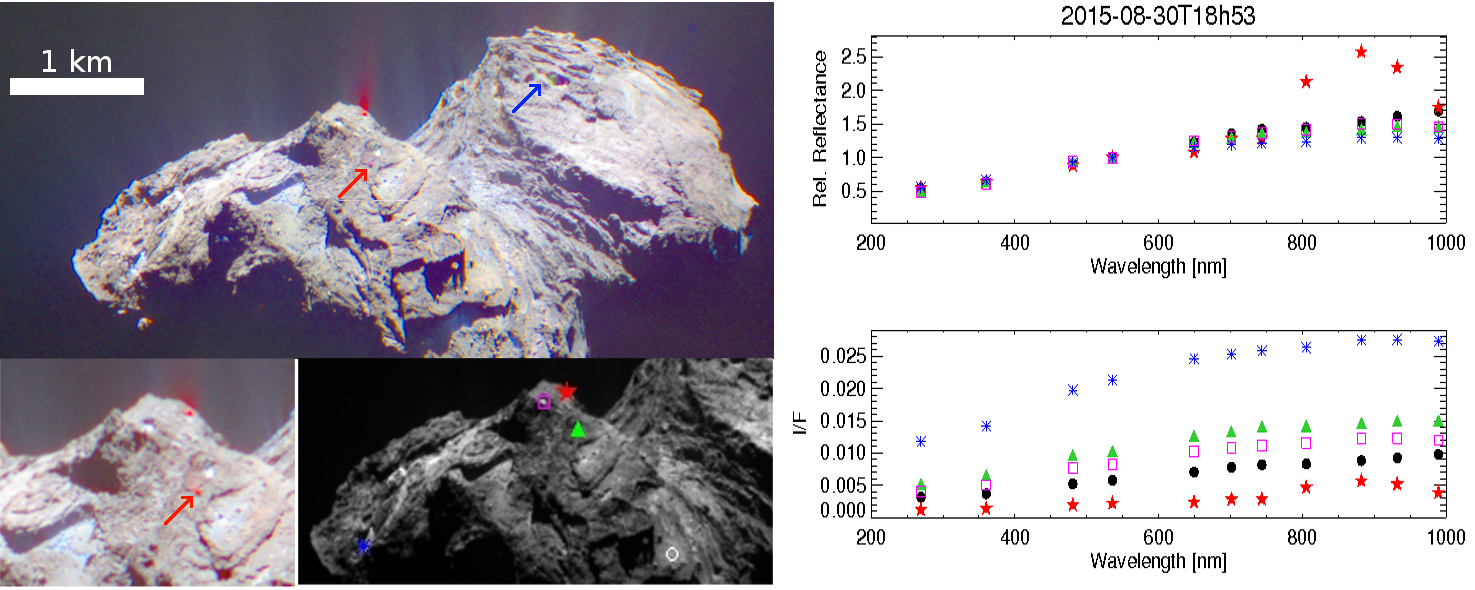}
\caption{Analysis of images obtained on 30 August 2015 at 18h51-18h54. Top left: RGB color image (composed from filters centered at 882, 649, and 480 nm) showing  a red jet (jet 123 in Table~\ref{all_jets}) departing from a bright spot, as well as  other faint jets that are indicated by the arrows. In particular, the cavity seen to be active at 6h49 on the same day still shows a very faint jet (jet 71 in Table~\ref{all_jets}) that is indicated by the red arrow. Bottom left:  Zoom into the region around the red jet in RGB colors produced with filters centered at 931, 649, and 480 nm; a faint jet departing from the 6h49 source can be visualized better, and the fainter flux of the red jet compared to the RGB in the top panel (the acquisition order of the filter was 269, 360, 743, 701, 480, 535, 649, 989, 931, 805, and 882 nm). Bottom center: image with symbols related to the five ROI.  Right panel: relative reflectance, normalized at 535 nm, vs. wavelength, and the reflectance at phase = 70$^{\circ}$ of five selected ROI (the red circle shows the dark terrain of the comet, the red star represents the red jet highlighted in the zoom, the blue asterisk shows a bright patch in Khonsu at which activity was previously observed, and the green triangle and magenta square indicate two different bright patches).}
\label{jet30aug_18h53}
\end{figure*}

Transient events with short lifetimes (shorter than two minutes) have been detected for the first time thanks to the unprecedented spatial and temporal coverage of the OSIRIS observations. The best example is a faint jet detected in the Bes region in images acquired on 30 August 2015, at UT 6h48-6h50 (Fig.~\ref{jet30aug_anal}, jet number 71 in Table~\ref{all_jets}). In this color sequence, the area hosting the jet (located precisely at longitude -140.6$^{\circ}$ and latitude -81.0$^{\circ}$, indicated by the yellow rectangle in Fig.~\ref{jet30aug} and by the red rectangle in Fig.~\ref{jet30aug_anal}) is not directly illuminated by the Sun at the time of the observations. This area is inactive in the first two images of the sequence (Fig.~\ref{jet30aug}), which were both acquired with the F15 filter centered at 269 nm. The activity then starts in the third image, and reaches its peak about 25 seconds later (image acquired with the F27 filter centered at 700 nm), after which the intensity progressively decreases, with almost no activity in the last two images of the sequence. We thus estimate its total duration to be about 95 s. As the flux changed during the sequence (which lasted for about 140 s), the jet spectrophotometry cannot be used to deduce information about the possible composition of the ejected material.  The jet is represented by the red asterisk in Fig.~\ref{jet30aug_anal}.  At its peak, the jet covers a projected area on the nucleus of about 20 pixels, corresponding to 1150 m$^2$. \\
Close to this jet, we observed a patch (represented by the blue asterisk in Fig.~\ref{jet30aug_anal}) that is 80\% brighter than the dark terrain (DT) of the comet. This patch is located in the Bes region at longitude -118.3$^{\circ}$ and latitude -81.8$^{\circ}$. Its spectrum is fainter than that of the comet DT. Previous studies of the nucleus of 67P have proven that the compositions of regions with this spectral behavior (i.e., relatively blue) include some water ice mixed with the comet DT (Pommerol et al., 2015; Fornasier et al., 2015, 2016, 2017; Barucci et al., 2016; Deshapriya et al., 2018; Filacchione et al., 2016a; Oklay et al., 2016, 2017). Two other bright patches are shown in Fig.~\ref{jet30aug_anal}, one in the Bes region (green triangle, longitude -119.2$^{\circ}$ and latitude -69.0$^{\circ}$), and one in the Khonsu region (magenta square, longitude -163.9$^{\circ}$ and latitude -13.2$^{\circ}$). The patch in Khonsu is three times brighter and spectrally flatter than the comet DT, indicating the exposure of some water ice. Its position corresponds to the location where several jets were identified in images obtained on 1 August 2015 (jet 138 in Table~\ref{all_jets}), showing activity for about three hours.This means that the water ice observed here was probalby freshly exposed after these activity events.

 The source area of jet 71 was repeatedly active during the perihelion passage as faint jets have been observed 12 times between June and October 2015 (see jet 71 in Table~\ref{all_jets} for more details). An example of the periodic activity is shown in Fig.~\ref{jet30aug_18h53}, where a very faint jet, indicated by the red arrow, is observed to depart from the same position as the short-lived event observed 12 hours before. The periodic nature of jets, that is, the exact same feature as was observed from one rotation to the next, has been noted and reported by Vincent et al. (2016a).  
Some other short-lived jets are observed in this image: a red jet in the Geb region (jet 123 in Table~\ref{all_jets}, indicated by the red star in Fig.~\ref{jet30aug_18h53} and in the zoom into the RGB image in the bottom left panel), whose activity starts in correspondence of observations with the F71 filter (989 nm), which reaches maximum in the last image of the sequence, acquired with the F41 filter (therefore it lasted longer than 25 seconds, and its duration is probably comparable to that of the jet at 6h49); and two faint jets, indicated by the blue arrow in Fig.~\ref{jet30aug_18h53}, departing from two cavities in the Wosret region, seen active periodically (cavities A and B, numbered 182 and 183 in Table~\ref{all_jets}). Again, we cannot constrain the potential composition of these jets  because of their short duration and flux variability in the different filters. The bright patches nearby the active regions (magenta square and green triangle in Fig.~\ref{jet30aug_18h53}) look slightly bluer in the near-infrared (NIR) region; this is consistent with the exposure of some water ice. The brightest spot is found in the Khonsu region (blue asterisk in Fig.~\ref{jet30aug_18h53}), at the same position as previously investigated in Fig.~\ref{jet30aug_anal}, showing that water ice exposed at the surface survives during the cometary day. Moreover, several exposures of water ice were later observed in this region, from January 2016, and they survived for several months (Deshapriya et al., 2016; Hasselmann et al., 2018).
\begin{figure*}
\centering
\includegraphics[width=0.95\textwidth,angle=0]{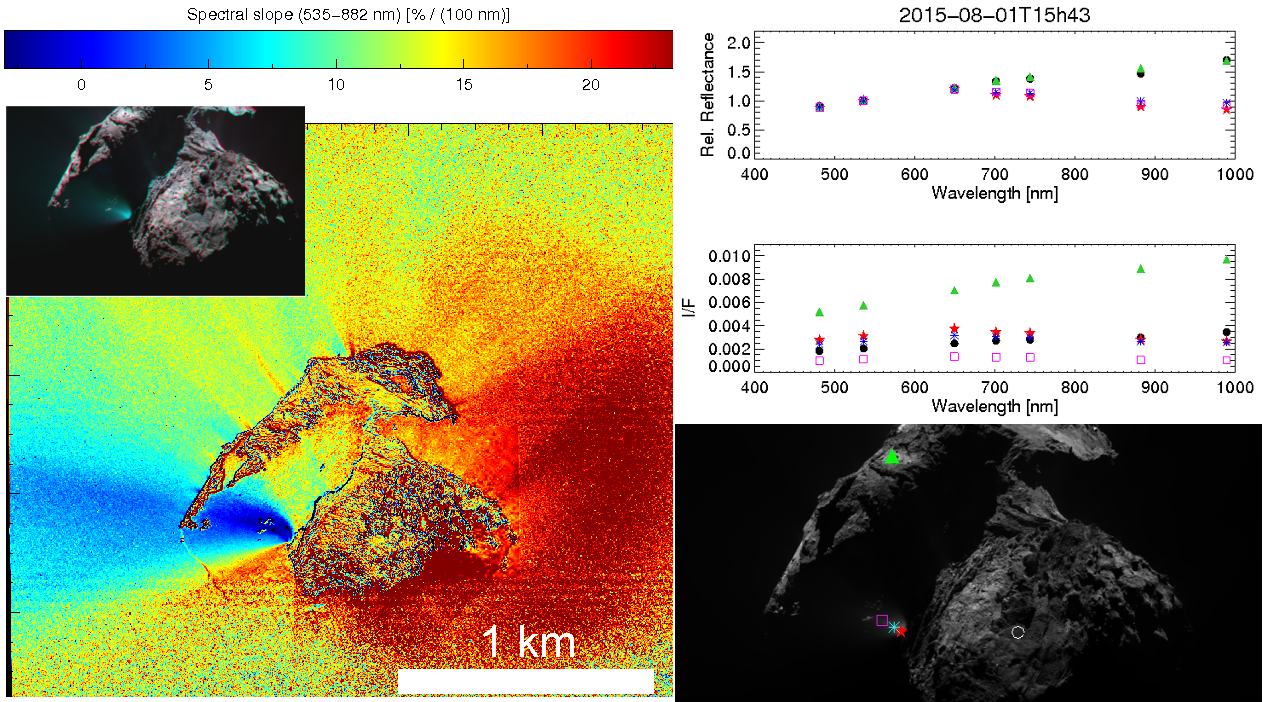}
\caption{Left: Slope of the spectra and RGB images from data obtained on 1 August acquired at 15h43, showing a jet in the shadows that departs from the Sobek region (jet 177 in Table~\ref{all_jets}). Right: Relative reflectance and I/F of the five ROI. The I/F of three positions on the jet (red star, cyan asterisk, and magenta square) is not corrected for the disk function because incidence and emission angle are not reliable in shadowed regions.}
\label{jet1aug}
\end{figure*}
\\
 Other examples of short-lived jets observed on 30 August 2015 are found in the Anhur region: two collimated jets that were reported in Fornasier et al. (2017; see Fig.~\ref{30aug2015}, panel at 12h21), which were active for about 50-70 s (jets 17 and 18 in Table~\ref{all_jets}), a jet observed at 8h09 that lasted for about 58 seconds (jet 13, Fig.~\ref{30aug2015}), and a jet observed at 22h34 that lasted for about 72 seconds (jet 20, Fig.~\ref{30aug2015}) and departed from an area in shadow. A few others on the same day are observed in Anuket, Bes, and Imhotep, and details are reported in Table~\ref{all_jets} (jets 36, 89, and 152). \\
Other short-lived jets are reported in Table~\ref{all_jets} and were observed on 1 August  (jet 74, Bes region), 26 July (jets 185 and 188, in Wosret, jet 4 in Anhur), 27 June (jet 182 in Wosret, jet 43 in Atum), and 11, 20, and 31 October 2015 (jets 98, 107, and 108 in Bes, jet 183 in Wosret). We note that some jets have blue, green, or red apparent colors because the activity was captured only with few filters in a color sequence, like the {\it red} jet in Geb we described before.

\subsection{Notable jets and outbursts} 
\begin{figure*}
\centering
\includegraphics[width=0.8\textwidth,angle=0]{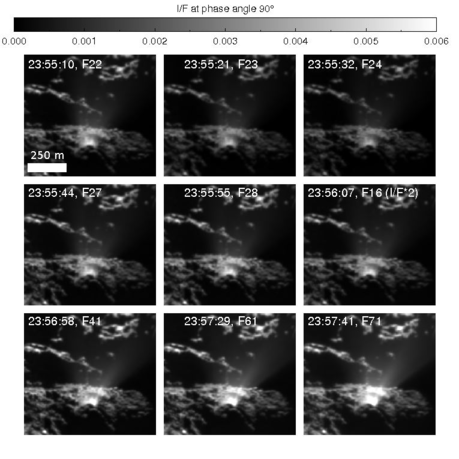}
\includegraphics[width=0.8\textwidth,angle=0]{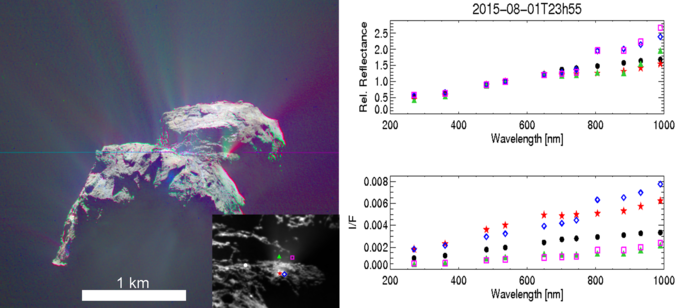}
\caption{Top: Nine of the 11 images of the sequence obtained on 1 August 2015 acquired at 23h55, showing the flux variation over time from a  jet with two close sources (jet 178 in Table~\ref{all_jets}). The I/F flux acquired with the F16 (centered at 360 nm) has been doubled to be correctly shown with the given intensity scale.
Bottom left: RGB images showing the double jet. The inset shows a zoom of the active region (in the F22 filter) with the five selected ROIs along the jet and on the comet nucleus (the white circle on the left side of the jet). Bottom right: Spectrophotometry and I/F of the five ROI (the nucleus DT is represented with a black circle). The horizontal line approximately in the middle of the RGB image is a residual of the combination of image subunits (an individual coregistration of three subregions of the full field of view was needed to improve the coregistration), and is and artifact.}
\label{jet1aug_double}
\end{figure*}
\begin{figure*}
\centering
\includegraphics[width=1.0\textwidth,angle=0]{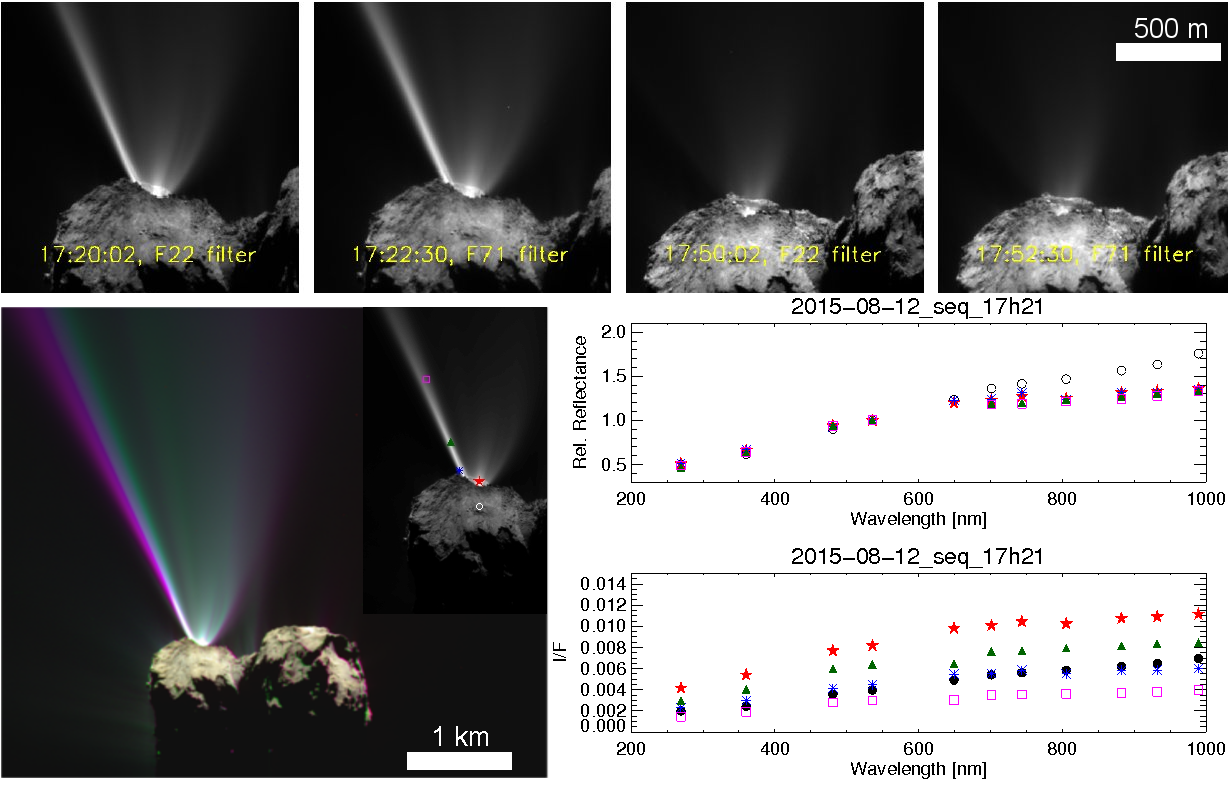}
\caption{Top: So-called perihelion outburst (jet 8 in Table~\ref{all_jets}) from two color sequences acquired on 12 August 2015 starting at 17h20 and 17h50. Bottom right: RGB color image (composed from filters centered at 882, 649, and 480 nm), and image of the selected ROI. The color difference at the left edge of the outburst in the RGB image arises because the comet rotated between the three exposures. Bottom left: Spectrophotometry of the nucleus and along the collimated component of the outburst.}
\label{perihelio}
\end{figure*}

\subsubsection{Two outbursts in Sobek}

Some of the observed jets or outbursts stand out particularly strongly. Two spectacular events from  1 August 2015 are shown in Figs.~\ref{jet1aug} and ~\ref{jet1aug_double} at 15h43 and 23h55 (jets 177 and 178 in Table~\ref{all_jets}), and both originate from the Sobek region. The 15h43 outburst has been reported by Vincent et al. (2016a) and displayed a broad plume (type B in their classification of transient events). Its intensity was about 11\% compared to that of the perihelion outburst of 12 August 2015, which is the brightest event reported in their list. These two outbursts took place some days after the discovery of a resolved boulder of about 0.8 m that orbited the comet and was observed at only 3.5 km from the Rosetta spacecraft (Fulle et al., 2016a, see their Fig.7). This indicates a progressive increase of the activity that is able to lift not only submillimeter and millimeter to centimeter grains from the surface, but also meter-sized boulders.  \\
Here we show for the first time the outburst colors and spectrophotometry (Fig.~\ref{jet1aug}), as well as the spectral slope of the nucleus and of the near coma. In contrast to the short-lived jets we discussed previously, these outbursts have a longer duration than that of the observing sequence, and we do not observe a fluctuation of the fluxes with time for the different filters. This means that their spectrophotometric properties are reliable. \\ 
Figure~\ref{jet1aug} clearly shows that the 15h43 outburst was spectrally bluer than the nucleus beyond 650 nm. Three areas in the very inner coma were monitored along the jet, and all show a negative slope in the 650-1000 nm range. This behavior may be attributed to grains that have an icy composition and/or small particle size.
On the surface of the comet, terrains with blue color beyond 650 nm have been observed in different regions, for example on Hapi (Fornasier et al., 2015). Infrared spectroscopy acquired with VIRTIS demonstrated this to be cometary dark material enriched in water ice (de Sanctis et al., 2015, Barucci et al., 2016). Icy particles in the ejecta are expected as the signature of water ice was reported in the July 2016 outburst (Agarwal et al., 2017), and the July 2015 cliff collapse in the Aswan region caused an outburst and exposed fresh icy material on the surface with an albedo beyond 40\% (Pajola et al., 2017).  \\
The spectrally blue color of the outburst may also be attributed to fine particles of micron or sub-micron size. With the OSIRIS data alone, we cannot constrain the grain size. However, we may deduce that the dominant size of the ejected particles is probably not much smaller than the incoming wavelength (i.e., $\ll$ 0.3-0.6 $\mu$m), otherwise we would have observed a higher flux and a negative spectral slope in the 260-600 nm range, as they would act as Rayleigh-type scatterers. \\
Other outbursts, such as the July 2016 dust plume, were also found to be composed of refractory and icy grains (Steffl et al. 2016; Agarwal et al. 2017), as the signature of icy particles was detected with the ALICE UV spectrometer. In particular, the models of the UV spectra acquired with ALICE permitted constraining the sizes of the grains: the icy component consisted of submicron grains, while refractories are ejected in larger grains with sizes of several hundred microns (Agarwal et al., 2017). Other outbursts have been observed by VIRTIS, for instance, like the one occurring on 13-14 September 2015 that was reported in Bockelee-Morvan et al. (2017). They measured large bolometric albedos, which indicates bright grains in the ejecta that are of silicatic or icy composition. The authors also measured a negative spectral slope in the IR region for the outburst. A slope like this is associated with high color temperatures (up to 630 K) of the ejected material, which the authors attributed to very fine dust particles (Bockelee-Morvan et al., 2017). We therefore conclude that the outburst we present here may include some icy grains mixed with dust particles, and/or grains of relatively small size, but larger than $\sim$ 0.2 $\mu$m.

Another spectacular event originating in the Sobek region is the double-component jet shown in Fig.~\ref{jet1aug_double} (jet 178 in Table~\ref{all_jets}), which was located at longitude 26.5$^{o}$ and latitude -14.7$^{o}$, at the boundary between the Sobek and Bastet regions. Individual images and the RGB composite image show that the jet has two individual sources: one that is constantly active during the whole sequence (the left side of the jet, indicated by the red star in Fig.~\ref{jet1aug_double}), which lasted for longer than 150 s, and one that became active at the end of the sequence, when images were acquired with the four filters covering the 800-1000 nm range. This source thus appears red, as underlined by the spectrophotometry of the blue symbol and magenta square in Fig.~\ref{jet1aug_double}. The component of the jet that was continuously active during the sequence is spectrophotometrically similar to the nucleus DT up to 650 nm, then it has a flatter behavior at longer wavelengths. Similarly as for the outburst described above, we interpret this as due to small icy grains mixed with dust.

\subsubsection{Perihelion outburst from Anhur}
\begin{figure*}
\centering
\includegraphics[width=1.0\textwidth,angle=0]{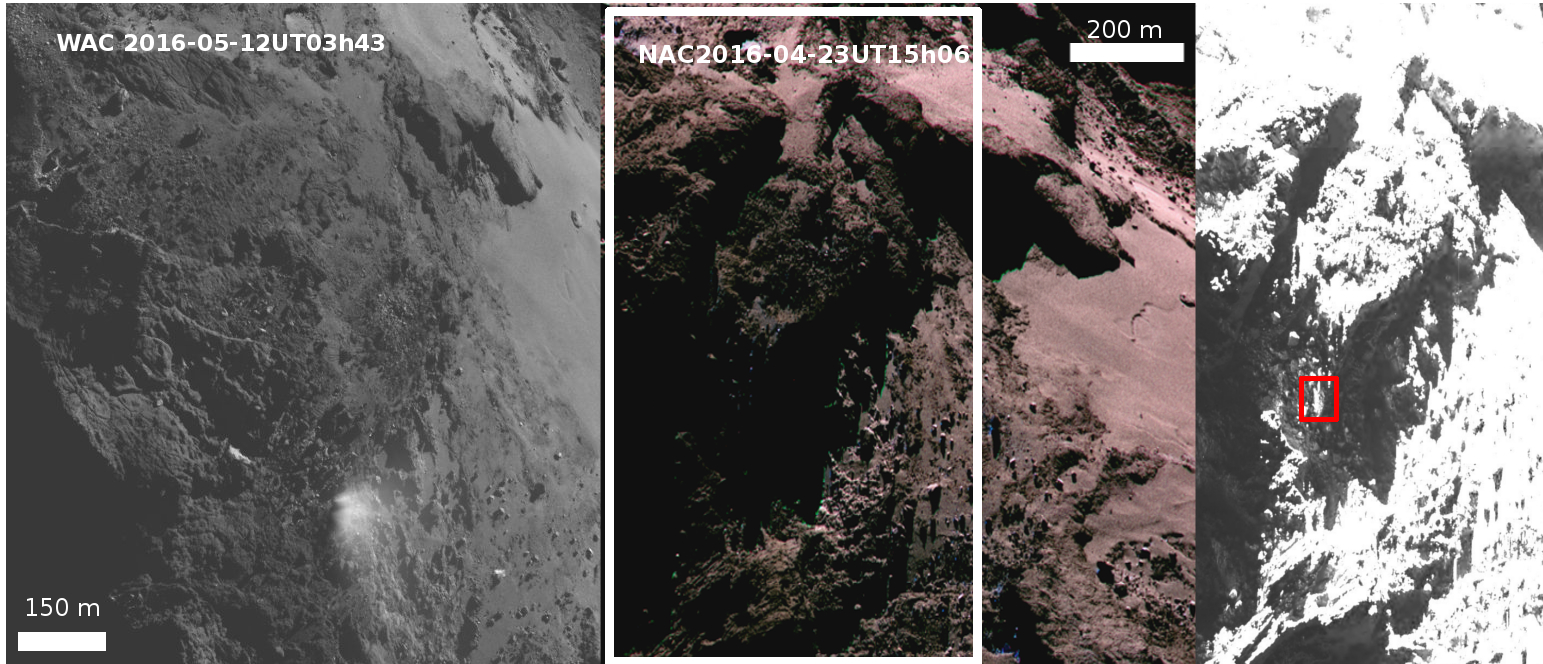}
\caption{Left panel: WAC image acquired on 12 May 2016 at UT 03:43 with the F12 filter centered at 630 nm showing an outburst in the Bes region, at the boundary with Imhotep (jets 114 and 115 in Table~\ref{all_jets}). Middle panel: RGB image composed from filters centered at 882, 649, and 480 nm from NAC images acquired on 23 April 2016. Right panel: Orange filter image corresponding to the area inside the white rectangle of the 23 April 2016 observations. Contrast is adapted to show details within the shadowed areas. The red rectangle indicates the source area of the jets observed on 12 May 2016.}
\label{jet_desh}
\end{figure*}
A similar spectral behavior is also observed for the spectacular outburst of 12 August 2015 (jet 8 in Table~\ref{all_jets}, Fig.~\ref{perihelio}). This is called the {\it \textup{perihelion outburst}} as it occurred a few hours before the comet reached perihelion. It was first reported by Vincent et al. (2016a), who estimated a total luminosity of 1.18$\times$10$^{13}$ W/nm at 649.2 nm and an ejected mass on the order of $\sim$ 100 tons, and who deduced that the source lies in the Anhur region. Lin et al. (2017) estimated for this event a mass ejection rate of $>$ 19 kg/s.  Here we present the spectrophotometry of the ejecta from the color sequence acquired around UTC 17:21, that is, close to the activity peak. The outburst has a complex shape, including a narrow jet and a complex structure. Very faint activity was reported in images acquired at UTC 17:06 and 18:06 with the orange F22 filter alone (Vincent et al. 2016a). The four selected ROI at the source and along the collimated jet all show a spectrally flat behavior beyond 650 nm, which we attribute to a mixture of dust and icy grains.

\subsubsection{Resolved outburst in Bes from May 2016 data}
\begin{figure*}
\centering
\includegraphics[width=0.9\textwidth,angle=0]{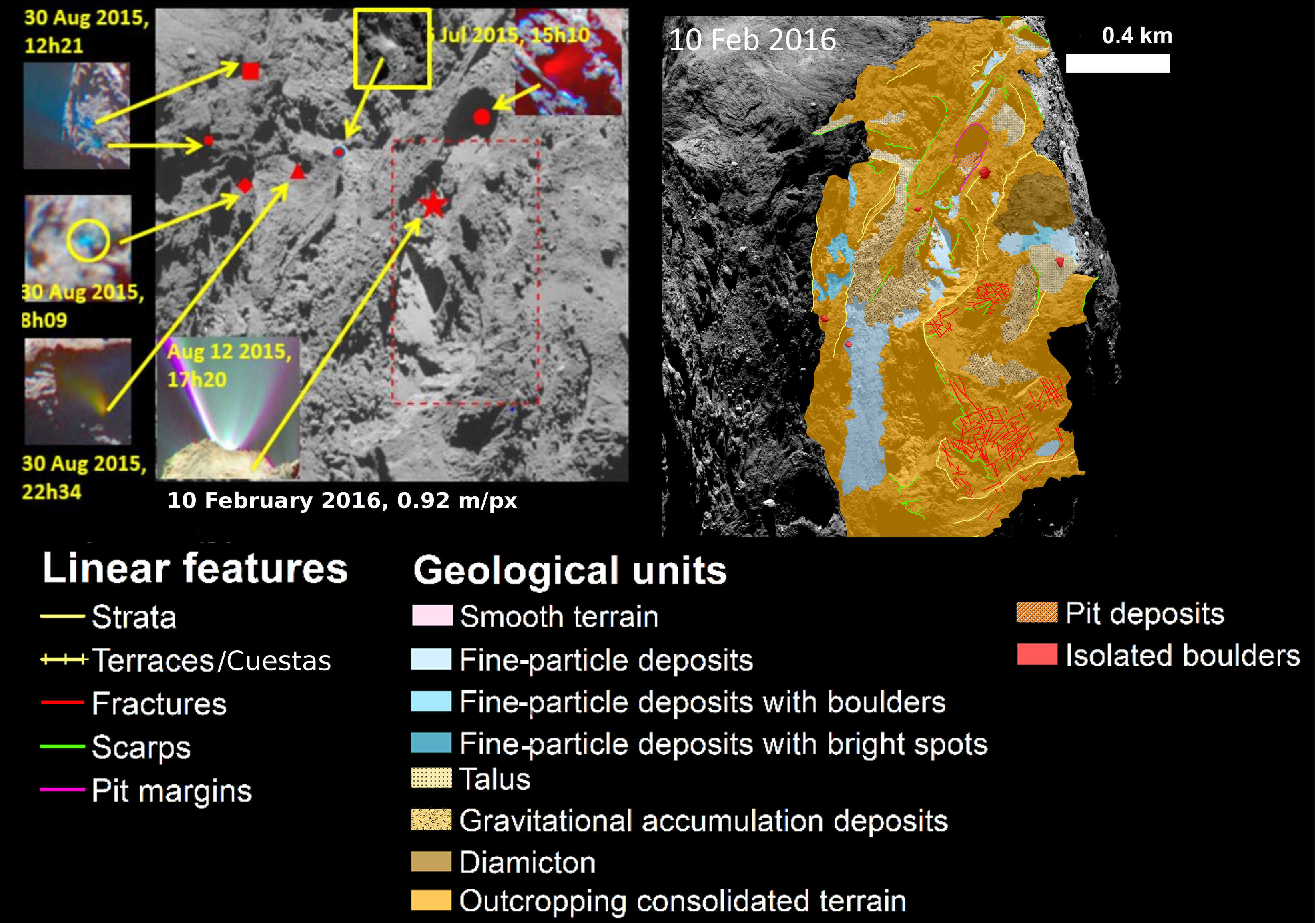}
\caption{Left panel:  Anhur region as seen on 10$^{}$ February 2016, UT 7h14. The locations of the jets identified close to perihelion passage are superimposed. Resolution: 92 cm/px. The symbols represent the corresponding positions of several jets: the star indicates the perihelion outburst, and the box indicates the uncertainties in the position of the jet sources; the circle shows a transient event on 26 July 2015, 15h10 (jet 4 in Table~\ref{all_jets}); squares show the double type-C outbursts on 30 August 2015 at 12h21 (jets 17 and 18 in Table~\ref{all_jets}), and the large square corresponds to the brighter of the two jets (jet 18); and the diamond indicates a transient event on 30 August 2015 at 8h09 (jet 13 in Table~\ref{all_jets}). Right panel: Geomorphological maps of the Anhur region from Fornasier et al. (2017).}
\label{Anhur}
\end{figure*}
Most of the events observed close to perihelion look faint compared to the outbursts reported by Vincent et al. (2016a), for instance. As the sequences were devoted to characterizing the nucleus, the exposure time was short, and several of the faint events have a low signal-to-noise ratio. Moreover, for safety reasons, Rosetta was far away from the nucleus, and the spatial resolution was therefore relatively poor (several m/px). During the last months of observations of  67P (i.e., May-September 2016), Rosetta came closer and closer to the nucleus, and caught a few outbursts at high resolution. This provided a glimpse of how the faint jets observed at perihelion would have looked like had Rosetta been closer to the nucleus.  A nice example is the 3 July 2016 outburst, which departied from an area of the Imhotep region. This was observed simultaneously with several Rosetta instruments (see Agarwal et al. (2017) for a detailed study of this event). 

Another resolved outburst, departing from the Bes region at the boundary with Imhotep, appeared in a single WAC image acquired on 12 May 2016 at 03h43 (jets 114 and 115 in Table~\ref{all_jets}, and Fig.~\ref{jet_desh}, left panel). This image was acquired when the comet was at an heliocentric distance of 2.98 AU, at a phase angle of 96.7$^{o}$, and with Rosetta at an altitude of 8.24 km, resulting in a spatial resolution of 0.82 m/px with the WAC camera. The outburst departs from two closely located sources at longitude [131.32$^{o}$,132.83$^{o}$] and at latitude [-65.62$^{o}$,-64.89$^{o}$]. Unfortunately, this region was always in the shadow during the high-resolution observations acquired with OSIRIS. Using the derived coordinates, we located the source of the jets in an NAC color sequence acquired on 23 April 2016, at UT15h06, when Rosetta was 28.5 km from the nucleus surface. This resulted in a spatial resolution of 0.55 m/px (Fig.~\ref{jet_desh}, middle panel). The geometry of the observations was different, the phase angle was higher (113$^{o}$), and the source of the jets was in shadow. However, the high dynamic range of the OSIRIS cameras allow us to study the morphology of the terrain within the shadowed regions, increasing the contrast. The source of the plume is located close to  scarp of about 40 m (Fig.~\ref{jet_desh}, right panel).\\
 The surface brightness of the plume was about 3.5 times higher than the surrounding regions of the nucleus, indicating the presence of bright material, likely icy grains. Within the shadow cast by the plume, the impinging light is attenuated by up to 50\%. This implies that the plume is optically thick, with an estimated optical depth of $\sim$ 0.65.

We calculated the filling factor $f$ , that is, the fraction
of radiance scattered by an optically thin dust coma,
to obtain the instantaneous total dust mass. The average radiance
factor $\overline{RADF}$ of the dust cloud was integrated in a circle with radius of 50 pixels, and subtracted by the
average surface radiance factor calculated inside an annulus external to
the event. The integrated dust filling factor in a aperture of $\pi R_{pixel}^{2}$
is expressed by (Knollenberg et al., 2016){\small \par}

\begin{equation}
f=\frac{\overline{RADF}}{w_{\lambda}(p_{(g)}/p_{(0)})}
,\end{equation}

where $w_{\lambda}$ is the single-scattering albedo from
Fornasier et al. (2015), estimated from Hapke modeling (Hapke, 2012)
of the surface scattering curve, $p_{(g)}$ is the particle phase
function from Bertini et al. (2017), and $p(0)$ is the extrapolation of this phase function for phase $g$ = 0. Assuming a differential grain
size distribution $n_{(r)}$, described by a power law $\propto r^{h+1}$,
and assuming spherical grains,  the expression for the integrated dust
mass is

\begin{equation}
M=2\rho\cdot f\cdot\pi R^{2}\frac{\intop n_{(r)}r^{2}dr}{\intop n_{(r)}rdr}
,\end{equation}

where $R$ is the aperture radius in meters and $r$
is the grain radius. As the power-law index $h$ is generally unknown,
we applied the indices estimated by Agarwal et al. (2017) from 10 $\mu m$ to 1 mm grain size for the multi-instrumental study of the 2016 July outburst. Therefore, from 10
to 150 $\mu m,$ we integrated Eq. 4 with $h=-2.54$, from 150
to 500 $\mu m$ with $h=-3.0$, and from 500 $\mu m$ up to 1 mm with $h=-6.9$ to
finally derive the mass density associated with the observed surface brightness. \\
We thus obtained a filling factor of 0.256, and an ejected dust mass for the given image in the range of 700-2220 kg for a grain bulk density $\rho$ ranging from 250 to 795\ $kg/m^{3}$ (Fulle et al., 2016b). For comparison, Agarwal et al. (2017) estimated an equivalent mass of 920$\pm$530 kg (in a given image, and for the same density range) for the July 2016 outburst, and a total ejected mass of 6500-118000 kg for a duration of between 14 and 68 min.  \\
Estimating the total mass ejected by the jets reported here is beyond scope of this paper, and moreover, the effective duration of most of the jets is unknown. However, it seems reasonable to assume that on average, the majority of the observed jets at perihelion, excluding the outbursts, ejected an instantaneous mass on the same order as the May 2016 event reported previously, that is, about one to a few thousand kilograms. 

\section{Morphology of active areas}

In this section we  describe the morphology of some of the southern hemisphere regions (Anhur, Sobek, Khonsu, Bes, and Wosret) that have been found to host several sources of the jets reported in this study.

\subsection{Anhur}

The Anhur region, which is illuminated for a relatively short interval during the cometary orbit, close to the perihelion passage, experiences strong thermal effects that result in a high degree of erosion.  Anhur is also highly active, and it is the source of several jets, as reported here and in previous papers (Vincent et al., 2016a, Fornasier et al., 2017).

\begin{figure*}
\centering
\includegraphics[width=0.8\textwidth,angle=0]{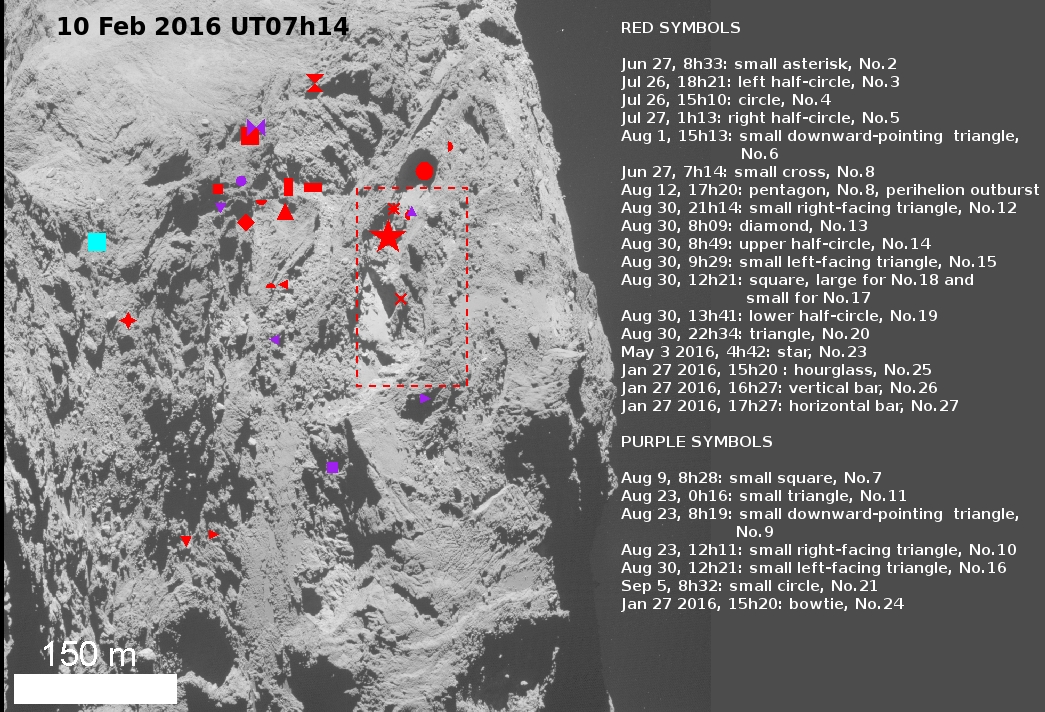}
\caption{Anhur region seen at high spatial resolution (92 cm/px) on 10 February 2016 at 7h14. The locations of the 26 individual activity sources identified close to perihelion passage are superimposed, as well as a few on 2016.}
\label{Anhur_alljets}
\end{figure*}

 This study identified 26 distinct jets originating from the Anhur region during July-October 2015, as reported in Figs.~\ref{Anhur} and ~\ref{Anhur_alljets}, and in Table~\ref{all_jets}. \\
The perihelion outburst source is not as precisely located as the others jets (see Figs.~\ref{Anhur} and ~\ref{Anhur_alljets}, red star and rectangle in a dashed red line), and our best estimate of its position is located inside a canyon-like structure with a pit and fine-particle deposits, where several exposures of water ice have been reported (Fornasier et al., 2017). Boulders and different types of deposits can be seen on the strata within or close to the jet site (Fig.~\ref{Anhur_alljets}), which likely originated by transport from steep slopes like the nearby cliffs (El-Maarry et al. 2015; Pajola et al. 2015; Lee et al., 2017), or from in situ degradation (Lee et al., 2017) due to sublimation and gravitational falls. 
\begin{figure*}
\centering
\includegraphics[width=0.75\textwidth,angle=0]{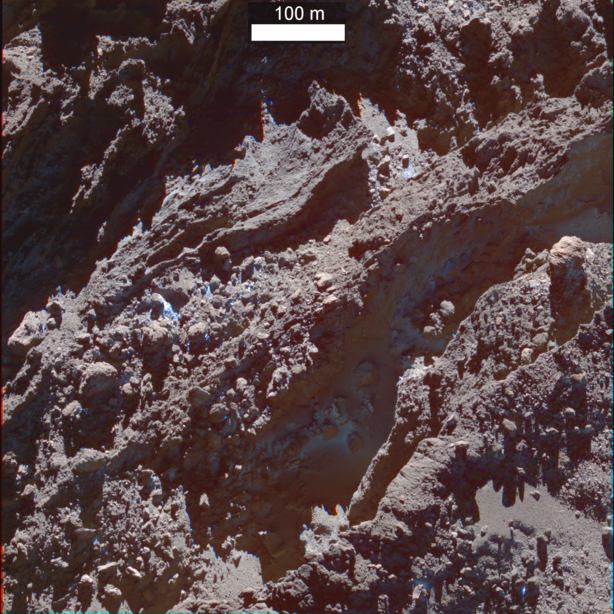}
\caption{RGB images in false colors of the Anhur region acquired on 25 June 2016 at UT 11:50, from a spacecraft altitude of 17.9 km, and a resolution of 35 cm/px. The color images are produced using the filters centered at 480 nm, 649 nm, and 882 nm. The heliocentric distance of the comet was 3.27 au.}
\label{Anhur_frost}
\end{figure*}

The sites of the 30 August 2015 transient events originating from the Anhur region are located on stratified terrain, at the base of the scarps (Fig.~\ref{Anhur_alljets}). The sources of the transient Anhur events at 12h21 (jets 17 and 18 in Table ~\ref{all_jets}, see Figs.~\ref{Anhur} and ~\ref{Anhur_alljets}) correspond to a talus formed at the base of a scarp, at the boundary with Khepry region, as well as to the 22h34 jet (jet 20 in Table ~\ref{all_jets}, see Fig.~\ref{30aug2015}). However, jets are found also on smooth terrains with fine-particle deposits, like the 8h09 event (Fig.~\ref{Anhur_alljets}, jet13 in Table ~\ref{all_jets}), or the optically thick plume observed on 27 January 2016 that was reported by Fornasier et al. (2017). Anhur also has an active pit, where a jet was observed on 26$^{}$ July 2015 (jet 4 in Table ~\ref{all_jets}, Fig.~\ref{Anhur_alljets}), as well as in previous observations from June 2015 reported in Fornasier et al. (2017).

The Anhur region is characterized by elongated canyon-like depressions in which cliffs expose sequences of strata. This indicates a pervasive layering. Local degradations and scarp retreats provide different types of deposits that partly cover flat tops, terraces, and bottoms of these depressions (Fig.~\ref{Anhur}). In this region, two bright patches of about 1500 m$^2$ each were observed on a flat terrace: one in a smooth terrain in Anhur, and the other just nearby, inside the Bes region, at the boundary with Anhur (Fornasier et al., 2016, 2017). These bright patches were observed between 27 April 2015 and 2-7 May 2015, and they lasted for at least ten days. Spectral modeling indicated a water ice abundance of 20-30\% mixed with the comet DT, which corresponds to a solid ice equivalent thickness of 1.5-27 mm (Fornasier et al., 2016). A few weeks before the detection of these exposed water ice patches, VIRTIS reported on 21-23 March 2015  the first and unique detection of CO$_2$ ice at the location of a patch entirely within the Anhur region (Filacchione et al., 2016b). In addition to these extended bright patches, evidence of exposure of volatiles was observed in the Anhur region in several instances (Fornasier et al., 2017). This region is thus characterized by compositional heterogeneities on the scale of tens of meters and volatile stratification, which gives rise to a particularly fragile terrain. A new scarp, 150 m long and 10 m high, formed at the boundary of the Anhur and Bes regions around perihelion passage or shortly after, in correspondence to the location of an extended bright patch (Fornasier et al., 2017). 

A close inspection of RGB color images of the Anhur region in 2016 reveals that water ice has been exposed in different locations in the form of tiny patches, or as frost hidden in the shadows cast by the canyon-like structure (Fig.~\ref{Anhur_frost}). Within the elongated depressions of Anhur, the deep strata of the large lobe of 67P, which are presumably enriched in volatiles, are exposed (Lee et al., 2017). An example of volatile frost and exposure of water ice close to a boulder is shown in Fig.~\ref{Anhur_frost} from 25$^{}$ June 2016 at UT 11h50. This observation was acquired at 35 cm/px resolution, and at a comet heliocentric distance of 3.27 au outbound. This spectacular view clearly indicates frost formation inside a shadowed region in Anhur, as well as exposure of ice, confirming that this region of the nucleus is one of the most highly enriched in volatiles, which is also proven by the high level of detected activity.

\subsection{Sobek}

\begin{figure*}
\centering
\includegraphics[width=0.9\textwidth,angle=0]{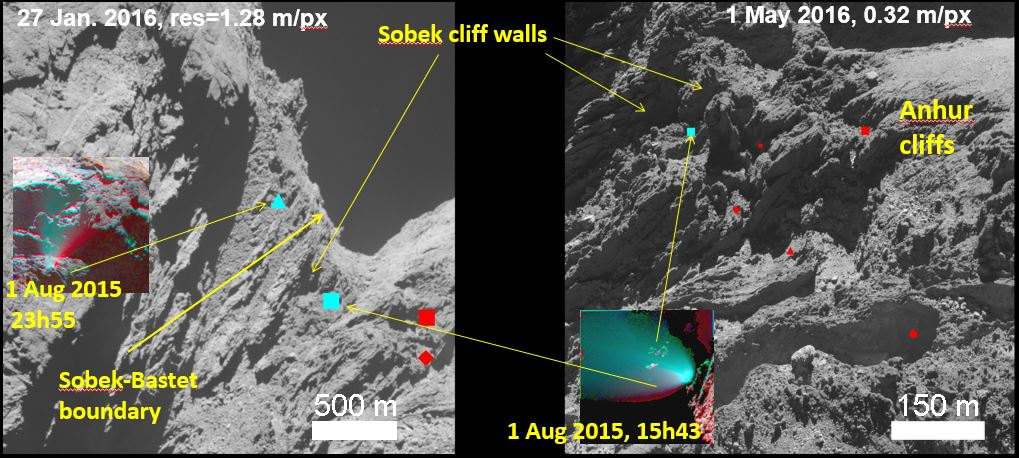}
\caption{Sobek region seen on 27 January 2016 at 18h20 with a spatial resolution of 1.28 m/px (on the left), and on 1 May 2016, 18h11, with a resolution of 0.32 m/px (on the right). The cyan square and triangle represent the two Sobek outburst locations as identified on 1 August 2015 at 15h43 and 23h55, respectively (jets 177 and 178 in Table~\ref{all_jets}), while the red symbols denote some jets that departed from the nearby Anhur region (see Fig.~\ref{Anhur}).}
\label{Sobek}
\end{figure*}

The Sobek region is located at the intersection between the large and small lobes in the southern hemisphere. In contrast to the Hapi region, which is also located between the two lobes but in the northern hemisphere, Sobek does not have widespread fine-particle deposits. However, this region has a low gravitational potential, and it shows an agglomeration of boulders as well as localized fine deposits in its central area (Lee  et al., 2017). The two spectacular outbursts observed on 1$^{}$ August 2015 (jets 177 and 178 in Table~\ref{all_jets}, Figs~\ref{jet1aug} and ~\ref{jet1aug_double}) are located at the foot of the scarp that separates the Sobek from the Bastet and Hapi regions. The source of the 15h43 outburst is located on a gravitational accumulation deposit and is surrounded by two highly vertical cliff walls on one side and the Anhur cliffs on the opposite side (see Fig.~\ref{Sobek}). The 23h55 event emerges  from an outcropping consolidated terrain. A total of six jets, including the two outbursts described above, originated from the Sobek region. 

\subsection{Khonsu}
\begin{figure}
\centering
\includegraphics[width=0.48\textwidth,angle=0]{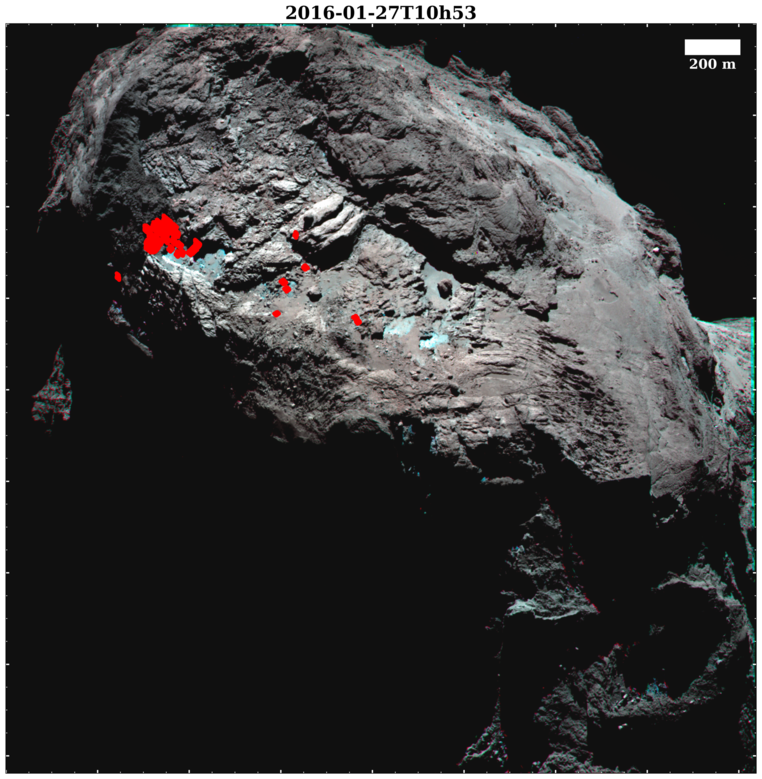}
\caption{Khonsu region seen at high spatial resolution (1.28 m/px) on 27 January 2016 at 10h53. The identified jets are superimposed in red. The cluster of red points is related to the jet that was periodically active (jet 138 in Table~\ref{all_jets}). A full description of the jet properties and associated morphological changes is reported in Hasselmann et al. (2018).}
\label{Khonsu}
\end{figure}
The Khonsu region, bounded by the Apis mesa, is dominated by outcropping consolidated terrain overlaid by some patches of fine-particle deposits, and it includes large boulders, a peculiar 200 m wide structure with three plate-shaped stacked features called the pancake feature, and evidence for layering (El-Maarry et al., 2016; Lee et al., 2017; Ferrari et al., 2018, see Fig.~\ref{Khonsu}). 

\begin{figure*}
\centering
\includegraphics[width=0.95\textwidth,angle=0]{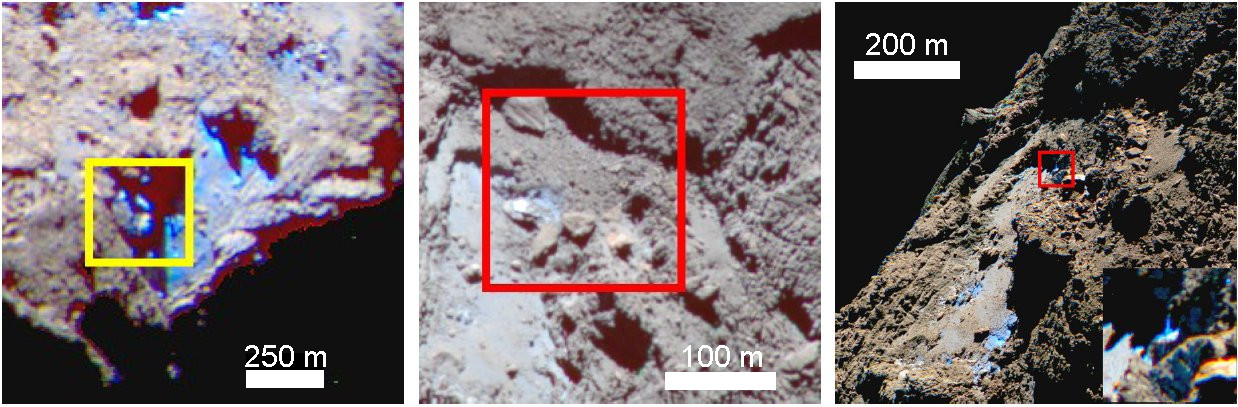}
\caption{Images showing, at different resolution, bright and relatively blue patches in or near the source of the 1 August 2015 events in Khonsu (jet 138 in Table~\ref{all_jets}):  26 July 2015 at 08h48 (left), 28 January 2016 at 01h48 (middle), and 2 July 2016 at 07h57 (right). The rectangles indicate the source region of the 1 August 2015 events (jet 138 in Table~\ref{all_jets}). }
\label{khonsu_zoom}
\end{figure*}

Several jets reported in Table~\ref{all_jets} originated from Khonsu (jets 137-146, see Fig.~\ref{Khonsu}). Two bright outbursts originated from source 138 (on 1$^{}$ August at 10h51 and 21h55-22h55, called event 7 by Vincent et al., 2016a), about one rotation apart. In particular, the source showed activity for about three hours since 10h51.  Many faint jets were also observed on the same day. The source of these outbursts is found on a rugged slope, at the foot of a cliff (see Figs.~\ref{Khonsu} and ~\ref{khonsu_zoom}), above a flat dust bank where bright patches have been spotted at least six days before the outburst (examples in Fig.~\ref{khonsu_zoom}), suggesting that this area is relatively abundant in water. Shortly after perihelion passage, exposures of water ice were detected, as discussed previously and shown in Figs. ~\ref{jet30aug_anal} and ~\ref{jet30aug_18h53}. The other jets reported in Fig.~\ref{Khonsu} originated mostly from fine-particle deposits. \\
 Hasselmann et al. (2018) reported several morphological changes that were all connected to this active area: the appearance of three ice patches, the formation of three shallow cavities, the sublimation of two thick dust layers, and the appearance of a 50 m jumping boulder that moved there from a nearby region. Previously, El-Maarry et al. (2017) also reported a 30 m boulder rolling into the southern reach of the same dust bank, while Deshapriya et al. (2016) noted another boulder hosting a spot rich in water ice that lasted for about half a year. All these morphological changes took place during the southern hemisphere summer.

\subsection{Bes}

\begin{figure}
\centering
\includegraphics[width=0.47\textwidth,angle=0]{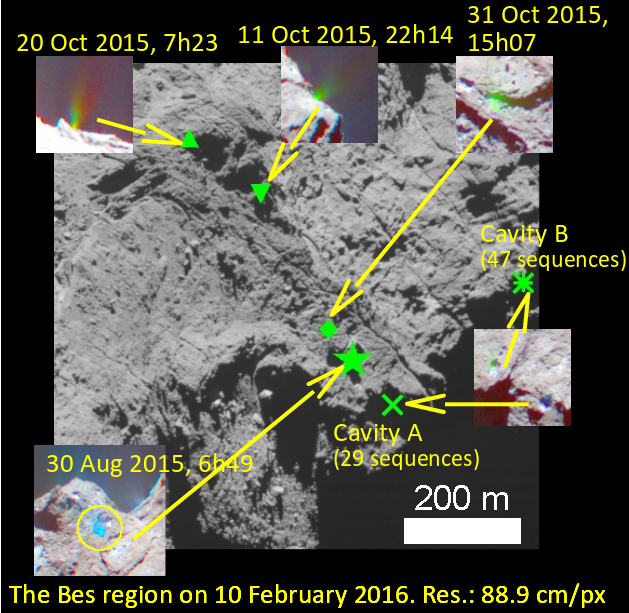}
\caption{Part of the Bes region seen at high spatial resolution (89 cm/px) on 10 February 2016 at 15h28. Some of the identified jets are superimposed. The green star represents the $\sim$95-second short-lived jet identified on 30 August 2015, at 6h49 (jet 71 in Table~\ref{all_jets}). The other symbols denote different transient events:  the diamond shows the event on 31 October,  15h07, the upward-pointing triangle shows the event on 20 October at 7h23, and the downward-pointing triangle represents the event on 11 October at 22h41 (listed in Table~\ref{all_jets} as jets 108, 107, and 98, respectively).}
\label{Bes}
\end{figure}

The Bes region is dominated by outcrops of consolidated terrain covered with deposits of fine materials. Diamictons and gravitational accumulation deposits are also found in the region (see Fig. 9 of Lee et al., 2017). In the onion-like layering of the nucleus (Massironi et al. 2015, Penasa et al., 2018), Bes is located in a shallower structural level than Anhur, and it also has mesa, high-slope ($> 35^{o}$) regions, and is sculpted by staircase terraces. \\
Many transient events (55, see Table~\ref{all_jets}) originated within this region (Fig.~\ref{Bes}), including the short-lived bright plume at 6h49 on 30 August 2015 (jet 71 in Table~\ref{all_jets}), and the resolved plume observed in the image in May 2016 (Fig.~\ref{jet_desh}, jets 114 and 115 in Table~\ref{all_jets}).
Some of these events arose from active cavities or alcoves located below a cliff, in a terrain with pervasive fractures (see Fig.~\ref{Bes}):\\
$\bullet$ Cavity A: longitude -119.1$\pm$3.9$^{\circ}$, latitude -80.1$\pm$4.1$^{\circ}$, seen active in 29 sequences (jet 111 in Table~\ref{all_jets})\\
$\bullet$ Cavity B: longitude -119.1$\pm$3.9$^{\circ}$, latitude -68.7$\pm$2.5$^{\circ}$, seen active in 47 sequences (jet 112 in Table~\ref{all_jets})\\
$\bullet$ Cavity C: longitude -111.5$\pm$3.3$^{\circ}$, latitude -69.8$\pm$3.5$^{\circ}$, seen active in 10 sequences (jet 113 in Table~\ref{all_jets})\\
These cavities very likely host volatiles because localized bright patches of water ice were observed in August 2015 (Figs.~\ref{jet30aug_anal}, and ~\ref{jet30aug_18h53}). 
 Long, linear fractures are visible across the cliff. They are probably formed by mechanical and/or thermal stresses. Small boulders are scattered at the surface below the cliff, while larger boulders are visible below an arc-shaped scarp (Fig.~\ref{Bes}).\\
In addition to cavities, the jet sources identified on Bes are also found  on dust deposits, at the foot of scarps and cliffs, and on consolidated terrains.

\subsection{Wosret}

Wosret is a region on the small lobe of 67P, and it is dominated by outcropping consolidated terrain with pervasive
fracturing. Fine-particle and gravitational accumulation deposits are observed together with several scarps and terraces arranged in staircase patterns that connect different strata (Lee et al., 2017). In Wosret, 33 sources were active in July-October 2015. In particular, several cavities or alcoves, that is, structures that cast shadows, were found to be repeatedly active close to perihelion, as shown in Fig.~\ref{Wosret}:
\begin{itemize}
\item Cavity A : located at longitude -28.7$\pm$3.5$^{\circ}$ and latitude -26.6$\pm$1.6$^{\circ}$, seen active in 61 color sequences (jet 182 in Table~\ref{all_jets}).
\item Cluster of B cavities : two or three closely placed cavities located at longitude -34.8
$\pm$3.6$^{\circ}$ and latitude -30.3$\pm$4.1$^{\circ}$, seen active in 33 color sequences (jet 183 in Table~\ref{all_jets})
\item Cavity C: located at longitude -40.7$\pm$1.6$^{\circ}$ and latitude -31.8$\pm$3.6$^{\circ}$, seen active in
16 sequences (jet 184 in Table~\ref{all_jets}).
\item Cavity D: located at longitude -14.7$\pm$3.1$^{\circ}$ and latitude -25.2$\pm$1.3$^{\circ}$, seen active in 37 color sequences (jet 185 in Table~\ref{all_jets})
\item Cavity E: located at longitude -15.3$\pm$2.8$^{\circ}$ and latitude -36.4$\pm$0.9$^{\circ}$, less active than the others cavities, as only three activity events were observed (jet 186 in Table~\ref{all_jets}).
\end{itemize}
The Wosret cavities usually emit faint cometary jets, sometimes with a very peculiar morphology, as highlighted in the insets on the left side of Fig.~\ref{Wosret}. These last events were brighter than the jets that were usually detected in these cavities, and occurred on 31 October, 20h49 (from cavity A, jet 182 in Table~\ref{all_jets}) and 21h49 (from cavity B, jet 183 in Table~\ref{all_jets}), with a collimated and broad shape.

These cavities, except for cavity A, were never observed at high spatial resolution and under good illumination conditions  during the mission. Tiny water ice patches were observed in cavity A. 

%


\section{Discussion}

This study demonstrates that several faint outbursts continuously contribute to the cometary activity at perihelion. They vary in duration and sometimes are extremely short (shorter than a few minutes). \\
Vincent et al. (2016b) reported that jets in the northern hemisphere arise mainly from rough terrains rather than smooth areas, and more specifically, from fractured walls. However, smooth areas also produce jets and jet-like features, as claimed by Shi et al. (2016, 2018). In the southern hemisphere, the jets and outbursts investigated here originated from both consolidated terrains (i.e., from scarps and cavities) and from smooth dust deposits that can sustain large boulders or fill niches or pit bottoms.
 
Several processes are evoked to explain the activity events reported here for comet 67P, and, more in general, for cometary nuclei:
\begin{figure*}
\centering
\includegraphics[width=0.9\textwidth,angle=0]{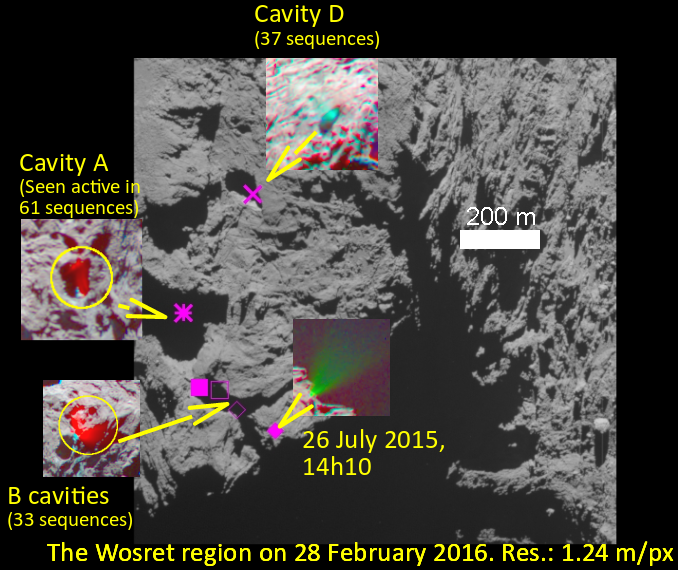}
\caption{Wosret region seen at a spatial resolution of 1.28 m/px on 28 January 2016 at 05h33. The identified jets are superimposed. The symbols represent different locations: the cross indicates cavity D (jet 185 in Table~\ref{all_jets}),
the asterisk shows cavity A (jet 182 in Table~\ref{all_jets}), the square represents the B cavities (jet 183 in Table~\ref{all_jets}), and the diamond indicates a
transient event seen in images of 26 July 2015 at 14h10 (jet 188 in Table~\ref{all_jets}).  }
\label{Wosret}
\end{figure*}
\begin{enumerate}
\item The main driver that causes jets is insolation coupled with local subsurface volatile enrichment and/or direct exposure of water ice at the nucleus surface. Owing to the complex morphology, different locations on the nucleus of 67P have varying diurnal illumination cycles. A study of the sublimation of water ice upon local sunrise is reported in Shi et al. (2018). 
The sublimation of water ice through the porous mantle, which Belton et al. (2010) defined as type I jets, is the likely driver of sources seen to be periodically active on comet 67P (e.g., the cavities listed in Table~\ref{all_jets} (jets 111-113 and 182-186), or the periodic jets reported in Vincent et al., 2016a). \\
The most active areas are those close to regional boundaries, which mostly correspond to cliff walls, or the insides of cavities or alcoves. These structures cast shadows that permit the recondensation of volatiles that arose from the subsurface during the cometary night, and partially from inner coma molecules that are backscattered to the nucleus surface (Davidsson \& Skorov, 2004; Crifo, 1987; Liao et al., 2018).  Subsurface thermal lag (Shi et al., 2016), coupled with the low thermal inertia of the comet, results in the recondensation of volatiles at night or in terrains that are often covered by shadows. \\
Moreover, as the comet approached perihelion, its dust mantle became thinner (Fornasier et al., 2016), exposing the underlying layers that are enriched in volatiles, and producing seasonal and diurnal color variations. Several instances of diurnal color changes and frost formation close to shadows have been observed. They were attributed to volatile recondensation during the cometary night (Fornasier et al., 2016, De Sanctis et al., 2015). Even far from perihelion, the complex local morphology coupled with the seasonal thermal lag permits local volatile recondensation, as shown in the Anhur region (Fig.~\ref{Anhur_frost}) at 3.3 au outbound. \\
Laboratory experiments on cometary analog mixtures  have shown that a considerable fraction of sublimating ice can be redeposited at the surface instead of being released through the dust mantle (Sears et al., 1999). They have also shown how material stratification, separation of types of ices, and the release of trapped gases could occur near the surface  of a comet, as has been observed on the nucleus of comet 67P (Fornasier et al., 2016, 2017; Filacchione et al., 2016b; Pajola et al., 2017).

\item Episodic and explosive events may be associated with cliff collapse (Vincent et al., 2016b). An example is the July 2015 outburst reported in Pajola et al. (2017), which exposed an inner layer, enriched in volatiles, in which bright and bluer material survived for several months. Several jets presented in this study were found below or close to cliffs or scarps, and some of them may be potentially related to a past cliff collapse. For instance, the sources of most of the Khonsu and Anhur jets were rich in volatiles and contained a number of scattered boulders that may have originated from a past cliff collapse that exposed volatile-rich layers. Unfortunately, most of the southern hemisphere was not observable before March 2015 from Rosetta, and therefore we do not have high-resolution images before perihelion to investigate the morphological changes associated with potential cliff collapses in detail. However, we noted the formation of a new 140 m long scarp nearby the boundary of Bes ad Anhur, as reported in Fornasier et al. (2017) and previously detailed in section 4.1, where exposure of ices and volatile stratification was reported before and after its formation (Fornasier et al., 2016, 2017; Filacchione et al., 2016b). Several jets have sources close to that scarp, but none of the events we reported here were directly associated with this cliff collapse and scarp formation, which took place sometime between August and December 2015.
  
\item Thermal stress produced the fractures that are ubiquitous on the surface of 67P. Fractures allow the heat wave 
to penetrate underlaying volatile-rich strata and may be the sources of jets, as discussed in Belton (2010) and in Bruck Syal et al. (2013). The jets located in an area of Bes characterized by long fractures (Fig.~\ref{Bes}) may be an example of this mechanism.

\item Local jets and outbursts such as the May 2016 event (jets 114 and 115 in Table~\ref{all_jets}, Fig.~\ref{jet_desh}) or the July 2016 dust plume (Agarwal et al., 2017) may have been produced by a pressurized reservoir of volatiles below the surface (Knollenberg et al., 2016). The exothermic transition of water ice from the amorphous to the crystalline state following sudden exposure to sunlight can cause the volatile outflow and trigger an activity event. According to Agarwal et al. (2017), this mechanism can take place even near the surface. 

\item Sinkhole collapse is invoked as the source of active pits (Vincent et al., 2015), which would result in the exposure of fresh volatiles on the cavity wall and interior after the collapse. However,  very few pits are observed in the southern hemisphere, and only one is seen to be active in the Anhur region (see Fornasier et al., 2017, and Fig.~\ref{Anhur}). 
The paucity of pits in the southern compared to the northern hemisphere is probably related to its higher insolation and erosion rate.

\end{enumerate}

Activity events on 67P are well localized on the southern hemisphere close to perihelion passage. In a similar manner, other comets observed by space missions showed activity departing from well-defined sources and not from the surface of the entire nucleus: only 10\% of the surface of 1P/Halley was estimated to be active (Keller et al., 1986) during the Giotto observations, while the hyperactive comet Hartley 2 showed well-localized jets mostly originating from the ends of its elongated nucleus (A'Hearn et al., 2011) and a plume of icy grains above the smooth waist (Protopapa et al., 2014).
Together with solar illumination, local compositional inhomogeneities are related to activity events. Several jet or outburst sources are located in or close to areas that are brighter and have colors that are relatively bluer than the dark terrain of the comet, indicating a local enrichment in volatiles that, once illuminated, sublimate. It has been noted that comet 67P shows evidence of local heterogeneities in composition at different spatial scales. Three types of terrains, from the spectrally bluer and water-ice-enriched terrains to the redder ones, associated mostly with dusty regions, have been identified by visible spectrophotometry with OSIRIS (Fornasier et al., 2015). The southern hemisphere shows a lack of spectrally red regions compared to the northern hemisphere. This is associated with the absence of widespread smooth or dust-covered terrains (Fornasier et al., 2016). \\ 
Local color and compositional heterogeneities have been identified from the OSIRIS images up to the decimeter scale (Feller et al., 2016) during the closest comet flyby on 14 February 2015, with  bright material, dark boulders, and some striae. During the 10$^{}$ April 2016 flyby, several bright spots associated with exposure of water ice mixed with the dark terrain of the comet  were reported (Feller et al., 2018; Hasselmann et al., 2017).  Bright patches showing exposure of water ice mixed with the dark terrain of the comet have been reported in several regions of the comet. Water ice amounts varied in these regions from a few percent (De Sanctis et al., 2015; Filacchione et al., 2016a; Pommerol et al., 2015; Barucci et al. 2016; Oklay et al., 2016) to $>$ 20\% in localized areas in the Anhur, Bes, Khonsu, and Imhotep regions (Fornasier et al., 2016, 2017; Deshapriya et al., 2016, Oklay et al., 2017), and at the Aswan site (Pajola et al., 2017). As reported in section 4.1, the Anhur region has shown local volatile stratification and wide ice patches before the southern spring equinox.  \\ 
Brightness variations in the comet surface at the centimeter and millimeter scale were reported by the CIVA instrument on board the Philae lander (Bibring et al., 2015), which observed a surface that was globally dominated by dark conglomerate, likely made of organics, with brighter spots that may be linked to mineral grains or point to ice-rich material. \\
Compositional inhomogeneities have also been reported on the surface of  comet 9P/Tempel 1 with the detection of dirty water-ice rich material (Sunshine et al., 2006), and they were associated with extensive subsurface sources of volatile material. Morphological changes were reported on comet Tempel 1 between the Deep Impact and the Stardust flybys, with fronts receding by several meters in a large smooth area (Veverka et al., 2013). Several jets were linked to the rugged surface bordering this smooth area (Farnham et al., 2007). These morphological changes were interpreted as the progressive sublimation and depletion of volatiles and ice-rich material (Meech et al., 2017). The short-period comets 9P/Tempel 1 and 81P/Wild 2, observed by the Deep Impact and Stardust missions, also showed extensive layering and stratification,  similar to comet 67P (Belton et al., 2007), and a complex morphology (Thomas et al., 2013).

\section{Conclusions}

We inspected over 2000 images acquired with the OSIRIS instrument on board Rosetta during four months around perihelion passage, and we identified and precisely located more than 200 transient events on the nucleus of 67P.\\
Our main findings are listed below.
\begin{itemize}
\item The source locations of the jets are usually found below cliffs, scarps, or inside cavities or alcoves that cast shadows, but they are also found on smooth terrains. Therefore, these activity events are not related to a specific terrain type or morphology, but are mainly driven by the local insolation.
\item This analysis indicates that several transient events observed at perihelion have lifetimes shorter than a few minutes. 
\item Faint jets are often periodic as a consequence of local illumination and of sublimation and recondensation processes of water ice. These processes, in particular, seem to be the source of the periodic jets that depart from cavities or alcoves.
\item Several jet sources are bright and spectrally bluer than the dark terrain of the comet. This implies a local enrichment of volatiles.
\item The ejecta of three outbursts we investigated have bluer colors in the visible-to-near-infrared range (but not in the near-ultraviolet region), indicating that the ejected material may contain some icy grains mixed with the ejected dust.
\item We reported a resolved bright plume observed in May 2016, which was optically thick. It had an instantaneous estimated mass loss of $\sim$ 1000-2000 kg. The faint jets observed at perihelion, whose durations are often unconstrained, probably eject a similar amount of material.
\end{itemize}
We presented a comprehensive inventory of source regions and locations of jets observed directly on the surface of 67P during and close to perihelion passage. This database of jets and outbursts can serve as a reference for further studies devoted to cometary activity, and, in particular, for future in situ space-probe observations of the activity and evolution of this comet, such as the NASA Caesar mission, if selected.

\vspace{0.3truecm}
\begin{acknowledgements}

OSIRIS was built by a consortium led by the Max-Planck-Institut f\"ur Sonnensystemforschung, Goettingen, Germany, in collaboration with CISAS, University of Padova, Italy, the Laboratoire d'Astrophysique de Marseille, France, the Instituto de Astrof\'isica de Andalucia, CSIC, Granada, Spain, the Scientific Support Office of the European Space Agency, Noordwijk, The Netherlands, the Instituto Nacional de T\'ecnica Aeroespacial, Madrid, Spain, the Universidad Polit\'echnica de Madrid, Spain, the Department of Physics and Astronomy of Uppsala University, Sweden, and the Institut  f\"ur Datentechnik und Kommunikationsnetze der Technischen Universitat  Braunschweig, Germany. \\ 
The support of the national funding agencies of Germany (DLR), France (CNES), Italy (ASI), Spain (MEC), Sweden (SNSB), and the ESA Technical Directorate is gratefully acknowledged.  We thank the Rosetta Science Ground Segment at ESAC, the Rosetta Mission Operations Centre at ESOC and the Rosetta Project at ESTEC for their outstanding work enabling the science return of the Rosetta Mission. SF acknowledges  the financial  support from the France Agence Nationale de la Recherche (programme Classy, ANR-17-CE31-0004). The authors thank  H. Campins for his comments that helped us improve this article.

\end{acknowledgements}

\newpage

\clearpage
\onecolumn

{\scriptsize
\begin{center}
\begin{longtable}{| c | c | c | >{\centering\arraybackslash}p{1.0cm} | c |c | >{\centering\arraybackslash}p{0.8cm} | p{5cm} |}
\caption[]{\noindent List of jet source positions as identified in June-October 2015 period, close to the perihelion passage, along with few jets sources identified in 2016 images. *Whether this jet originated from the stated cavity is uncertain; $^1$ Fornasier et al., 2017; $^2$ Agarwal et al., 2017. LT stands for local time (on a 24-hour basis). Repeat indicates the repeatability of a given jet in one location (usually 1 except for periodic jets).}
\label{all_jets} \\
\hline \multicolumn{1}{|c|} {\textbf{\# jet}} & \multicolumn{1}{c|}
{\textbf{Lon ($^\circ$)}} &\multicolumn{1}{c|}
{\textbf{Lat ($^\circ$)}} & \multicolumn{1}{>{\centering\arraybackslash}p{1.0cm}|} {\textbf{Region}} & \multicolumn{1}{c|}
{\textbf{Repeat}} & \multicolumn{1}{>{\centering\arraybackslash}p{0.8cm}|} {\textbf{Type}} & \multicolumn{1}{c|}
{\textbf{LT}} & \multicolumn{1}{p{5cm}|} {\textbf{Time $\&$ Description}} \\  \hline
\endfirsthead
\multicolumn{6}{c}{\tablename\ \thetable\ -- \textit{continued from previous page}} \\ \hline 
\multicolumn{1}{|c|} {\textbf{\# jet}} & \multicolumn{1}{c|}
{\textbf{Lon ($^\circ$)}} &\multicolumn{1}{c|}
{\textbf{Lat ($^\circ$)}} & \multicolumn{1}{>{\centering\arraybackslash}p{1.0cm}|} {\textbf{Region}} & \multicolumn{1}{c|}
{\textbf{Repeat}} & \multicolumn{1}{>{\centering\arraybackslash}p{0.8cm}|} {\textbf{Type}} & \multicolumn{1}{c|}
{\textbf{LT}} & \multicolumn{1}{p{5cm}|} {\textbf{Time $\&$ Description}} \\  
\hline 
\hline 
\endhead
\hline
\multicolumn{4}{c}{\tablename\ \thetable\ -- \textit{Continued on next page}} \\ 
\endfoot
\hline \hline
\endlastfoot
        1 & -53.4 & -12.7 & Aker & 1 & B & 14h20m &5 September 2015, 10h41. Faint.\\
\hline
        2 & 61.5 & -42.2 & Anhur & 1 & B & 16h05m & 27 June 2015, 8h33. Faint, found on limb.\\
\hline
        3  & 62.5 & -43.4 & Anhur & 1 & B & 15h07m &26 July 2015, 18h21. Faint, possible sublimation.\\
\hline
        4 & 65.5 $\pm$ 1.1 & -39.8 $\pm$ 0.1 & Anhur & 2 & B & 09h09m & 26 July 2015, 15h10. Clear jet but in shadows. T $\sim$ 62s.\\
           & & & & & & &  31 October 2015, 22h49. Possible ice sublimation.\\ 
\hline 
        5 & 69.3 & -39.0 & Anhur & 1 & B & 04h51m &27 July 2015, 1h13. Faint.\\
\hline
        6 & 4.1 & -56.9 & Anhur & 1 & B & 20h48m & 1 August 2015, 15h13. Faint.\\
\hline
        7 & 40.0 & -63.5 & Anhur & 1 & B & 00h25m & 9 August 2015, 8h58. Faint, found on limb.\\
\hline
        8 & 59.9 $\pm$ 5.8 & -51.5 $\pm$ 12.3 & Anhur & 5 & B, C & 14h11m &  An outburst (C) on 12 August 2015, 17h20. T $>$ 30 minutes (seen again at 17h50). Brightest event of the 2015 perihelion passage of comet 67P.\\
           & & & & & B & &  27 June 2015, 7h14 $\&$ 7h53. Faint, possible ice sublimation.\\
           & & & & & B & &  22 August 2015, 23h16 . Faint.\\
\hline
        9 & 44.4 & -32.7 & Anhur & 1 & B & 06h21m &23 August 2015, 8h19. Faint.\\
\hline
        10 & 54.3 & -61.0 & Anhur & 1 & B & 14h30m &23 August 2015, 12h11. Faint.\\
\hline
        11 & 63.2 & -43.1 & Anhur & 1 & B & 15h50m &23 August 2015, 0h16. Faint, found on limb.\\
        12 & 9.1 & -59.7 & Anhur & 1 & B & 21h24m & 30 August 2015, 21h14. Faint.\\
\hline
        13 & 46.4 & -35.2 & Anhur & 1 & B & 22h19m &30 August 2015, 8h09. Bright. T $\sim$ 58s.\\
\hline
        14 & 45.5 & -41.9 & Anhur & 1 & B & 23h31 & 30 August 2015, 8h49. Faint.\\
\hline
        15 & 47.2 & -42.5 & Anhur & 1 & B & 00h56m & 30 August 2015, 9h29. Faint.\\
\hline
        16 & 42.4 & -46.8 & Anhur & 1 & B & 06h14m &30 August 2015, 12h21. Faint.\\
\hline
        17 & 45.0 & -31.3 & Anhur & 1 & C & 06h26m &30 August 2015, 12h21. Faint. T $\sim$ 54s.\\
\hline
        18 & 51.2 & -29.2 & Anhur & 1 & C & 06h50m & 30 August 2015, 12h21. Bright. T $\sim$ 67s.\\
\hline
        19 & 49.7 & -34.4 & Anhur & 1 & B & 09h20m & 30 August 2015, 13h41. Faint.\\
\hline
        20 & 52.2 & -36.2 & Anhur & 1 & B & 02h55m & 30 August 2015, 22h34. Bright. T $\sim$ 72s.\\
\hline
        21 & 47.3 & -32.4 & Anhur & 1 & B & 16h55m & 5 September 2015, 8h32. Faint.\\
\hline
        22 & 63.5 & -33.8 & Anhur-Bes & 1 & B & 00h05m& 5 September 2015, 11h41. Faint, found on limb.\\ 
\hline  
        23 & 29.6 & -32.3 & Anhur & 1 & B & 09h23m &3 May 2016, 4h42. Faint.\\
\hline
        24 & $\sim$52 & $\sim$-29 & Anhur & 1 & B & 07h54m & 27 January 2016, 15h20. Bright outburst$^1$.\\
\hline
        25 & $\sim$59 & $\sim$-29 & Anhur & 1 & B & 08h18m & 27 January 2016, 15h20. Bright outburst$^1$.\\
\hline
        26 & 53.6 & -34.6 & Anhur & 1 & B & 10h14m & 27 January 2016, 16h27. Bright$^1$.\\
\hline
        27 & 56.4 & -35.7 & Anhur & 1 & B & 12h25m &27 January 2016, 17h27. Bright$^1$.\\
\hline

        28 & -70.7 & -7.5 & Anuket & 1 & B & 19h42m &1 August 2015, 17h13. Faint\\
\hline
        29 & -53.1 & -19.8 & Anuket & 1 & B & 07h00m & 1 August 2015, 22h25. Faint, found on limb.\\
\hline 
        30 & -70.2 & -27.5 & Anuket & 1 & B & 19h27m & 9 August 2015, 10h13. Faint, found on limb.\\
\hline
        31 & -78.8 & 7.8 & Anuket & 1 & B & 16h35m & 9 August 2015, 21h23. Faint.\\
\hline
        32 & -64.2 & -28.7 & Anuket & 1 & A & 20h11m & 9 August 2015, 22h43. Faint.\\
\hline
      33 & -64.8 & -18.6 & Anuket & 1 & B & 05h23m &22 August 2015, 23h16. Faint.\\
\hline
        34 & -56.7 & 3.6 & Anuket & 1 & B & 11h00m & 23 August 2015, 14h11. Faint.\\
\hline
        35 & -67.1 & -20.3 & Anuket & 1 & B &00h14m & 30 August 2015, 13h01. Faint.\\
\hline
        36 & -65.0 & -22.9 & Anuket & 1 & B &09h18m &30 August 2015, 17h33. Bright. T $>$ 12s.\\
\hline
        37 & -64.2 & -28.6 & Anuket & 1 & B & 03h03m &30 August 2015, 14h21. Faint.\\
\hline
        38 & -62.8 & -18.8 & Anuket & 1 & B & 01h30m &31 August 2015, 1h14. Faint.\\
\hline
        39 & -61.6 & -16.8 & Anuket & 1 & B & 22h29m &5 September 2015, 15h06. Faint.\\
\hline
        40 & 132.8 & 27.9 & Ash & 1 & B & 12h12m & 27 June 2015, 16h28. Faint, found on limb.\\
\hline
        41 & 120.6 & 28.9 & Ash & 1 & B & 12h35m & 27 June 2015, 17h08. Faint, found on limb.\\
\hline 
        42 & 110.6 & 13.4 & Ash & 1 & A & 14h28m & 1 August 2015, 8h18. Bright.\\
\hline
       43 & -127.2 & -17.6 & Atum & 1 & B & 03h14h & 27 June 2015, 20h50. Bright. T $\sim$ 100s.\\
\hline
        44 & -153.1 & 11.3 & Atum & 1 & B & 09h06m &26 July 2015, 10h18. Faint.\\
\hline
        45 & -135.8 & -24.1 & Atum & 1 & B & 19h42m &1 August 2015, 7h08. Faint.\\
\hline
        46 & -125.0 & -19.3 & Atum & 1 & B & 14h09m & 1 August 2015, 16h13. Faint.\\ 
\hline
        47 & -121.5 & -63.6 & Atum & 1 & B & 16h03m & 9 August 2015, 10h13. Faint, found on limb.\\
\hline
        48 & -143.7 & 5.2 & Atum & 1 & B &07h19m &12 August 2015, 8h28. Faint.\\
\hline
       49 & -125.3 & -33.6 & Atum & 1 & B & 01h17m & 22 August 2015, 23h16. Faint.\\
\hline
        50 & -139.6 & -30.3 & Atum & 1 & B & 02h20m & 23 August 2015, 0h16. Faint.\\
\hline
        51 & -125.5 & -19.8 & Atum & 1 & A & 13h26m & 30 August 2015, 9h29. Faint.\\
\hline
        52 & -144.2 & -13.9 & Atum & 1 & B & 04h00m &30 August 2015, 17h33. Faint. \\
\hline
        53 & -149.4 & 3.1 & Atum & 1 & B & 13h48m & 5 September 2015, 13h41. Faint.\\
\hline
        54 & -124.5 & -24.5 & Atum-Anubis & 1 & B & 15h30m & 5 September 2015, 13h41. Clear, in shadows.\\
\hline
        55 & 15.5 & -1.4 & Bastet & 1 & B & 13h01m & 1 August 2015, 10h51. Faint.\\
\hline
        56 & 8.9 & -11.1 & Bastet & 1 & B & 15h32m & 1 August 2015, 12h21. Faint.\\
\hline
        57 & 16.5 & -11.5 & Bastet & 1 & A & 02h28m & 1 August 2015, 17h43. Faint.\\
\hline
        58 & 18.3 & -9.8 & Bastet & 1 & B & 07h14m &30 August 2015, 13h41. Faint\\
\hline
        59 & 34.2 & -21.7 & Bastet-Sobek & 1 & B & 08h8m & 20 October 2015, 10h05. Faint.\\
\hline
        60 & 74.0 & -33.4 & Bes & 1 & B & 14h21m & 27 June 2015, 7h14. Faint, found on limb.\\
\hline
        61 & -127.5 $\pm$ 0.1 & -77.7 $\pm$ 0.4 & Bes & 2 & B &04h00m-03h24m &26 July 2015, 6h48 $\&$ 18h51. Faint.\\
\hline
        62 & -99.7 & -80.1 & Bes & 1 & C & 12h42m &26 July 2015, 10h18. Faint.\\
\hline
        63 & 82.5 & -29.7 & Bes & 1 & B & 00h48m &26 July 2015, 10h18. Faint.\\
\hline
        64 & -102.8 $\pm$ 0.1 & -80.3 $\pm$ 0.1 & Bes & 2 & B & 11h32m & 26 July 2015, 22h13. Faint.\\
           & & & & & & 15h22m & 11 October 2015, 14h41. Faint.\\
\hline
        65 & -125.7 & -67.7 & Bes & 1 & B & 19h22m & 26 July 2015, 14h40. Faint.\\
\hline
        66 & -125.1 & -81.6 & Bes & 1 & C & 00h41m & 26 July 2015, 17h21. Faint.\\
\hline
        67 & 149.1 & -81.5 & Bes & 1 & B & 20h49m & 26 July 2015, 18h21. Clear jet in shadows.\\
\hline
        68 & 78.7 & -32.5 & Bes & 1 & B & 20h00m &26 July 2015, 20h21. Bright, found on limb.\\
\hline
        69 & -105.6 & -79.8 & Bes & 1 & B &12h24m & 26 July 2015, 22h43. Faint.\\
\hline
        70 & -134.8 & -70.7 & Bes & 1 & B & 10h24m &26 July 2015, 22h43. Faint.\\
\hline
        71 & -142.7 $\pm$ 2.7 & -81.7 $\pm$ 1.1 & Bes & 12 & B &06h42m &  30 August 2015, 18h53. Faint. \\ 
           & & & & &                                      A, B &  &  26 July 2015, 23h43 (B). Faint.\\
          & & & & & & & 1 August 2015: 8h00 (B), 8h18 (B), 8h43 (B), 10h51 (B), 22h25 (B), 22h55 (B), 23h25 (A). All faint.\\
          & & & & & B & 11h21m&  23 August 2015, 15h11 . Faint.\\
          & & & & & B &        11h13m &  30 August 2015, 6h49 . Bright. T $\sim$ 95s.\\
          & & & & & B & 07h04m&  31 October 2015, 16h07 . Faint.\\
\hline
        72 & -151.2 & -75.3 & Bes & 1 & B & 22h35m & 1 August 2015, 9h08. Faint.\\
\hline
        73 & -129.1 & -71.2 & Bes & 1 & A & 07h18m &1 August 2015, 12h51. Faint.\\
\hline
        74 & 73.2 & -40.3 & Bes & 1 & B & 22h45m & 1 August 2015, 13h51. Clear jet in shadows. T $\sim$ 70s. \\
\hline
        75 & 85.8 & -45.5 & Bes & 1 & B & 19h07m & 1 August 2015, 23h55. Faint.\\
\hline
        76 & 105.7 & -61.3 & Bes & 1 & B & 18h28m &1 August 2015, 22h55. Faint.\\
\hline
        77 & 110.3 & -57.3 & Bes & 1 & B & 13h10m & 1 August 2015, 7h38. Faint.\\
\hline
        78 & -95.9 & -78.7 & Bes & 1 & B & 12h24m & 9 August 2015, 7h28. Faint, found on limb.\\
\hline
        79 & 87.3 $\pm$ 2.8 & -49.1 $\pm$ 0.4 & Bes & 3 & B & 12h59-15h35 & 9 August 2015, 13h50, 14h30 $\&$ 15h10. Faint, possible ice sublimation.\\
\hline
        80 & -99.0 & -78.6 & Bes & 1 & B &  12h01m & 9 August 2015, 19h42. Faint, found on limb.\\
\hline
        81 & 77.6 & -32.8 & Bes & 1 & B & 14h25m & 12 August 2015, 16h50. Faint.\\
\hline
        82 & 79.4 & -48.0 & Bes & 1 & B & 09h42m &12 August 2015, 14h22. Faint.\\
\hline
        83 & -89.5 & -82.6 & Bes & 1 & B & 01h44m &22 August 2015, 22h16. Faint, found on limb.\\
\hline
        84 & -130.9 & -64.4 & Bes & 1 & B & 05h36m & 23 August 2015, 1h38. Faint, found on limb.\\
\hline
        85 & -128.1 & -78.3 & Bes & 1 & A & 07h45m &23 August 2015, 2h38. Faint, found on limb.\\
\hline
        86 & -133.5 & -82.0 & Bes & 1 & A &09h18m & 23 August 2015, 3h38. Faint.\\
\hline
        87 & -73.2 & -79.7 & Bes & 1 & B & 20h30m & 23 August 2015, 7h19. Faint, found on limb.\\
\hline
        88 & -134.2 & -67.4 & Bes & 1 & B &13h12m & 30 August 2015, 21h54. Faint.\\
\hline
        89 & -104.4 & -67.4 & Bes & 1 & B &16h08m &30 August 2015, 10h09. Bright. T $>$ 12s.\\
\hline
        90 & 72.6 & -30.8 & Bes & 1 & B & 17h06m &30 August 2015, 16h53. Faint.\\
\hline
        91 & 78.3 & -31.0 & Bes & 1 & B & 02h00m & 30 August 2015, 21h14. Faint.\\
\hline
        92 & -131.0 & -66.7 & Bes & 1 & B & 21h48m &5 September 2015, 17h06. Faint, found on limb.\\
\hline
        93 & -108.3 & -78.1 & Bes & 1 & B & 14h19m &5 September 2015, 12h41. Faint, found on limb.\\
\hline
        94 & 109.7 & -79.1 & Bes & 1 & B & 17h06m& 5 September 2015, 6h32. Faint, near limb.\\
\hline
        95 & 130.1 & -70.4 & Bes & 1 & B & 20h28m & 5 September 2015, 7h32. Faint, found on limb.\\
\hline
        96 & 153.7 & -69.3 & Bes & 1 & B & 20h37m & 5 September 2015, 19h06. Faint.\\
\hline
        97 & 168.1 & -70.1 & Bes & 1 & B & 19h39m &5 September 2015, 18h06. Faint.\\
\hline
        98 & -179.3 & -72.2 & Bes & 1 & B & 23h40m & 11 October 2015, 22h14. Clear. T $\geq$ 12s.\\
\hline
        99 & 117.6 & -67.4 & Bes & 1 & B & 21h26m & 11 October 2015, 23h14. Faint.\\
\hline
        100 & 82.2 & -40.4 & Bes & 1 & B & 23h05m &12 October 2015, 1h14. Faint.\\
\hline
       101 & -178.2 & -75.8 & Bes & 1 & B & 13h17m & 19 October 2015, 19h20. Faint, found on limb.\\
\hline
        102 & -172.1 & -74.4 & Bes & 1 & A & 11h42m &19 October 2015, 18h20. Faint, found on limb.\\
\hline
        103 & -146.4 & -77.4 & Bes & 1 & B & 13h13m & 20 October 2015, 6h23. Faint, found on limb.\\
\hline
        104 & 73.6 & -53.9 & Bes & 1 & B & 15h11m & 20 October 2015, 12h05. Faint.\\
\hline
        105 & 88.2 $\pm$ 1.4 & -49.1 $\pm$ 0.3 & Bes & 2 & B & 10h15m & 20 October 2015, 9h05 $\&$ 10h05. Faint.\\
\hline
        106 & 91.2 & -61.4 & Bes & 1 & B & 14h24m & 20 October 2015, 11h05. Faint.\\
\hline
        107 & 167.8 & -67.7 & Bes & 1 & B & 12h12m & 20 October 2015, 7h23. Clear. T $\sim$ 46s.\\
\hline
        108 & -155.0 & -80.1 & Bes & 1 & B & 00h11m & 31 October 2015, 15h07. Clear. T $\sim$ 34 s.\\
\hline
        109 & 66.5 & -87.2 & Bes & 1 & B & 06h08m & 31 October 2015, 22h49. Faint.\\
\hline
        110 & 121.0 & -69.6 & Bes & 1 & B & 18h35m & 31 October 2015, 15h07. Faint.\\
\hline
        111 & -114.5 $\pm$ 7.5 & -80.1 $\pm$ 4.1 & Bes & 29 & A, B, C & & Active cavity A of the Bes region. A bright spot is frequently observed from the cavity around perihelion:\\
                & & & & & & & Faint events on 26 July 2015: 11h40 (A), 12h10 (B), 17h51* (B), 23h13 (B), 23h43 (C).\\
                & & & & & & & Faint type B events on 27 July 2015: 0h13, 0h43.\\
                & & & & & & & Faint type B events on 1 August 2015: 12h51, 13h51, 15h13, 15h43, 21h25, 22h25, 23h55.\\
                & & & & & & & Faint event (B) on 23 August 2015, 9h19. Found on limb.\\
& & & & & & 08h36m &  Faint type B events on 30 August 2015: 7h29, 8h49, 14h21*, 18h53, 19h33, 21h14, 21h54.\\           & & & & & & & Faint event (B) on 11 October 2015, 12h41.\\
                & & & & & & & Faint type B events on 31 October 2015: 11h46, 17h07*, 18h07*, 19h07, 20h49, 21h49.\\
\hline
        112 & -119.6 $\pm$ 5.0 & -68.5 $\pm$ 2.4 & Bes & 47 & A, B, C & & Active cavity B of the Bes region. A bright spot is frequently observed from the cavity around perihelion:\\
                & & & & & B & & Faint events on 26 July 2015: 11h40, 12h10, 12h40, 13h10, 13h40, 17h21.\\
                & & & & & & & Faint events on 27 July 2015: 0h13 (B), 0h43 (C), 1h13 (B).\\
                & & & & & & & Faint events on 1 August 2015: 10h51 (B), 12h51 (A), 15h13 (B), 15h43 (B), 16h43 (B), 17h43 (B), 18h13 (B), 21h55 (B), 23h25 (B), 23h55 (B).\\
                & & & & & B & 05h36m & Faint events on 30 August 2015: 9h29, 10h09, 11h41, 16h53, 17h33, 21h14, 21h54, 22h34, 23h54.\\
                & & & & & B & & Faint events on 23 August 2015: 7h19* (found on limb), 9h19.\\
                & & & & & B & & Faint events on 11 October 2015: 14h41*, 21h14.\\
                & & & & & B & & Faint event  on 19 October 2015: 19h20*.\\
                & & & & & B  & &  Faint events on 20 October 2015: 3h23*, 7h23*.\\
                & & & & & B & & Faint events on 31 October 2015: 10h46, 11h46, 12h46, 15h07, 16h07, 17h07, 18h07, 19h07, 20h49, 21h49, 22h49, 23h49.\\
\hline
        113 & -111.5 $\pm$ 3.3 & -69.8 $\pm$ 3.5 & Bes & 10 & B, C && Active cavity C of the Bes region:\\
                & & & & & & 18h44m & Faint type B events on 30 August 2015, 11h41 $\&$ 22h34*.\\
                & & & & & & & Faint event (B) on 5 September 2015, 11h41. Found on limb.\\
                & & & & & & & Faint event (B) on 19 October 2015, 23h02*\\
                & & & & & & & Faint event (B) on 31 October 2015, 10h46*.\\
                & & & & & & & Faint type B events on 20 October 2015: 0h02, 1h02, 12h05*.\\
                & & & & & & & Faint event (B) on 11 October 2015, 16h53*.\\
                & & & & & & & Faint event (C) on 9 August 2015, 7h13*. Found on limb.\\
\hline
        114 & 132.0 & -64.9 & Bes & 1 & B & 12h08m  & 12 May 2016, 3h43. “Twin” bright outburst along with $\#$115.\\
\hline
        115 & 133.8 & -65.3 & Bes & 1 & B & 12h14m & 12 May 2016, 3h43. “Twin” bright outburst along with $\#$114.\\
\hline
         116 & 15.3 & -83.6 & Geb & 1 & B & 15h27m &26 July 2015, 7h48. Faint.\\
\hline
         117 & -101.9 & -76.4 & Geb & 1 & B & 00h16m & 1 August 2015, 8h18. Faint.\\
\hline
        118 & -70.6 & -58.8 & Geb & 1 & B & 15h48m & 1 August 2015, 15h13. Faint\\
\hline
        119 & -99.3 & -53.8 & Geb & 1 & B & 13h40m & 9 August 2015, 8h13. Faint.\\
\hline
        120 & -95.4 & -59.2 & Geb & 1 & B & 17h45m & 9 August 2015, 10h13. Faint but clear jet.\\
\hline
        121 & -54.2 & -65.5 & Geb & 1 & B & 04h08m & 22 August 2015, 22h16. Faint.\\
\hline
        122 & -43.0 & -66.0 & Geb & 1 & A & 22h30m & 23 August 2015, 7h19. Faint but clear, found on limb.\\
\hline
        123 & -98.0 $\pm$ 0.7 & -67.5 $\pm$ 1.1 & Geb & 4 & B &00h45m &  23 August 2015, 10h19. Faint.\\
          & & & & & A & 09h42m & 30 August 2015: 18h53 (T $>$ 24s) \\
          & & & & & A &11h01m &        30 August 2015:19h33. Faint \\
          & & & & & B &16h55m &       30 August 2015:22h34. Faint\\
\hline
        124 & -94.9 & -77.1 & Geb & 1 & B & 19h50m & 30 August 2015, 11h41. Faint.\\
\hline
        125 & -66.5 & -72.7 & Geb & 1 & B &02h57m &30 August 2015, 14h21. Faint.\\
\hline
        126 & 19.5 & -80.0 & Geb & 1 & B & 05h42m &31 August 2015, 0h34. Faint.\\
\hline
        127 & -94.9 & -45.6 & Geb-Hapi & 1 & B & 13h35m &5 September 2015, 11h41. Faint.\\
\hline
        128 & -75.6 & -51.1 & Geb & 1 & B &16h51m &5 September 2015, 12h41. Faint.\\
        129 & 9.6 & -82.6 & Geb & 1 & B & 08h54m & 20 October 2015, 11h05. Faint.\\
\hline
        130 & 42.1 & -73.7 & Geb & 1 & B & 09h08m & 20 October 2015, 10h05. Faint.\\
\hline
        131 & -100.2 & -27.1 & Hapi & 1 & B & 10h52m & 5 September 2015, 22h45. Faint.\\
\hline
        132 & -20.3 & 4.5 & Hatmehit & 1 & B & 12h40m &12 August 2015, 7h00. Faint.\\
\hline
        133 & 96.8 & -21.1 & Khepry & 1 & B & 12h16m & 1 August 2015, 7h38. Faint.\\
\hline
        134 & 76.8 & -2.3 & Khepry & 1 & B & 19h10m &12 August 2015, 7h00. Faint.\\
\hline
        135 & 93.2 & -28.0 & Khepry & 1 & B & 21h06m & 30 August 2015, 18h13. Faint.\\
\hline
        136 & 79.1 & -25.9 & Khepry & 1 & B & 15h46m & 20 October 2015, 0h02. Faint.\\
\hline
        137 & -161.4 & -47.6 & Khonsu & 1 & B &04h00m &26 July 2015, 20h21. Faint.\\
\hline
        138 & -163.6 $\pm$ 2.1 & -12.2 $\pm$ 1.7 & Khonsu & 12 & A, B, C &  & 1 August 2015:\\
           & & & & & &16h44 &  A bright outburst at 10h51 (B) and lasted for 4 consecutive sequences: 11h21 (C), 11h51 (A), 12h21 (A), 12h51 (B).\\
           & & & & & &  &  Other sequences, all faint: 20h55 (C), 21h25 (B), 21h55 (B), 22h25 (B), 22h55 (B), 23h25 (B), 23h55 (A).\\
\hline
        139 & -161.0 $\pm$ 0.1 & -18.7 $\pm$ 0.1 &Khonsu & 2 & A & 17h42 & 12 August 2015, 13h52 $\&$ 14h22. Faint.\\
\hline
        140 & -169.0 & -26.3 & Khonsu & 1 & B & 02h18m & 30 August 2015, 17h33. Faint. \\
\hline

        141 & -126.5 & -64.5 & Khonsu-Bes & 1 & A & 20h08m & 5 September 2015, 16h06. Faint.\\
\hline
        142 & -163.5 & -44.4 & Khonsu-Imhotep & 1 & B & 12h07m  &20 October 2015, 6h23. Faint, found on limb.\\
\hline
        143 & -155.0 & -45.0 & Khonsu & 1 & B & 06h47m & 20 October 2015, 3h23. Faint.\\
\hline
        144 & -147.7 & -65.2 & Khonsu-Bes & 1 & B & 08h36m & 31 October 2015, 19h07. Faint.\\
\hline
        145 & -162.7 & -46.6 & Khonsu-Imhotep & 1 & B & 14h56m & 31 October 2015, 22h49. Faint.\\
\hline
        146 & -147.1 & -81.2 & Khonsu & 1 & B & 15h56m &31 October 2015, 22h49. Faint.\\
\hline
        147 & 157.8 & -12.1 & Imhotep & 1 & B & 18h23m & 1 August 2015, 8h43. Faint.\\
\hline
        148 & 154.6 & -7.4 & Imhotep & 1 & B & 18h59m &1 August 2015, 9h08. Faint.\\
\hline
        149 & 178.9 & -49.6 & Imhotep & 1 & B & 02h50m & 1 August 2015, 12h21. Faint.\\
\hline
        150 & -170.4 & -8.2 & Imhotep & 1 & B & 19h47m & 9 August 2015, 13h50. Faint, possible ice sublimation.\\
\hline
        151 & 137.0 & -8.2 & Imhotep & 1 & B & 15h23h &12 August 2015, 15h20. Faint.\\
\hline
        152 & -179.4 & -43.3 & Imhotep & 1 & B & 20h39m &30 August 2015, 15h01. Bright. T $>$ 24s. \\
\hline
        153 & 167.2 & -60.7 & Imhotep & 1 & B & 00h42m & 30 August 2015, 17h33. Faint.\\
\hline
        154 & 120.8 & -37.2 & Imhotep & 1 & B & 14h36m &5 September 2015, 17h06. Faint.\\
\hline
        155 & 131.8 & -36.9 & Imhotep & 1 & B & 18h36m & 5 September 2015, 6h32. Faint.\\
\hline
        156 & -166.6 & -46.8 & Imhotep & 1 & B & 17h15m &20 October 2015, 9h05. Faint, found on limb.\\
\hline
        157 & -159.6 & -53.6 & Imhotep & 1 & B & 17h45m & 20 October 2015, 9h05. Faint.\\
\hline
        158 & 172.0 & -33.2 & Imhotep & 1 & C & 10h35m &3 July 2016, 7h36. Bright outburst$^2$.\\
\hline
        159 & 14.4 & 25.6 & Ma'at & 1 & B &09h07m  & 27 June 2015, 6h33. Faint, found on limb.\\
\hline
        160 & 19.5 & 13.6 & Ma'at & 1 & B & 09h25m & 27 June 2015, 6h33. Faint, found on limb.\\
\hline
        161 & 6.8 & 13.9 & Ma'at & 1 & B & 11h12m & 27 June 2015, 7h53. Faint, found on limb.\\
\hline
        162 & 7.3 & 11.5 & Ma'at & 1 & B & 09h57m & 27 June 2015, 7h14. Faint, found on limb.\\
\hline
        163 & -23.6 & 10.2 & Ma'at & 1 & A & 11h25m & 27 June 2015, 21h30. Faint, found on limb.\\
\hline
        164 & -17.8 & 17.6 & Ma'at & 1 & B & 11h49m & 27 June 2015, 21h30. Faint, found on limb.\\
\hline
        165 & -17.0 & 15.0 & Ma'at & 1 & B & 11h55m & 27 June 2015, 21h30. Faint, found on limb.\\
\hline
        166 & -27.7 & -1.6 & Maftet & 1 & A & 12h31m & 27 June 2015, 22h10. Faint, found on limb.\\
\hline
        167 & -56.4 & -22.2 & Neith & 1 & A &  11h08m &1 August 2015, 12h21. Faint\\
\hline
        168 & 9.7 & -25.7 & Neith & 1 & B & 12h04m & 1 August 2015, 22h55. Faint.\\
\hline
        169 & -54.4 & -25.4 & Neith & 1 & B & 05h03m &30 August 2015, 15h01. Faint.\\
\hline
        170 & -47.2 & -29.7 & Neith & 1 & B & 10h30m& 30 August 2015, 17h33. Faint.\\
\hline
        171 & -27.4 & 14.8 $\pm$ 0.3 & Nut & 2 & B & 12h31m & 27 June 2015, 8h33 $\&$ 20h50. Faint, possible ice sublimation.\\
\hline
        172 & -25.7 & 12.2 & Nut & 1 & B & 12h37m &27 June 2015, 22h10. Faint, found on limb.\\
\hline
        173 & -41.9 & 14.2 & Serqet & 1 & B & 12h48m & 27 June 2015, 22h50. Faint, found on limb.\\
\hline
        174 & -53.2 & 19.3 & Serqet & 1 & B & 18h23m & 9 August 2015, 21h23. Faint.\\
\hline
        175 & -161.2 & 68.3 & Seth & 1 & B & 07h40m  & 19 July 2015, 4h36. Faint.\\
\hline
        176 & -178.7 & 77.9 & Seth & 1 & B & 06h28m & 19 July 2015, 4h36. Faint.\\
\hline
        177 & 34.0 & -23.6 & Sobek & 1 & B & 23h47m & 1 August 2015, 15h43. An outburst. T $\geq$ 141s.\\
\hline
        178 & 26.5 & -14.7 & Sobek & 1 & C & 15h13m & 1 August 2015, 23h55. An outburst. T $\geq$ 151s.\\
\hline
        179 & -2.7 & -43.0 & Sobek & 1 & A & 01h12m &23 August 2015, 7h19. Faint but clear.\\ 
\hline
        180 & -56.9 & -39.8 & Sobek & 1 & B & 22h13m &30 August 2015, 23h54. Faint but clear jet.\\
\hline
        181 & -71.5 & -35.5 & Sobek-Anuket & 1 & B & 09h39m &31 October 2015, 17h07. Faint.\\
\hline
        182 & -28.7 $\pm$ 3.5 & -26.6 $\pm$ 1.6 & Wosret & 61 & A, B & & Active cavity A of the Wosret region:\\
        & & & & & & &  Faint events on 26 July 2015: 6h48 (B), 18h21 (B), 18h51 (B), 19h21 (A).\\
& & & & & A & 19h26m &  Bright event (A) on 26 July 2015, 23h43. T $\sim$ 150s. \\
                & & & & & B & &  Faint events on 27 July 2015: 0h43, 1h13.\\    
        & & & & & A & 20h25m &  Bright event (A) on 27 July 2015, 0h13. T $\sim$ 150s. \\ 
                & & & & & &  18h36m &  Faint events on 1 August 2015: 6h38 (B), 7h38 (B), 8h00 (B), 8h18 (B), 11h51 (B), 12h21 (B), 12h51 (B), 13h21 (B), 13h51 (B), 16h13 (B), 16h43 (B), 17h43 (A), 18h13 (B), 18h43 (B), 20h55 (B), 21h25 (B), 21h55(B).\\
                & & & & & B & 01h25m & Bright event (B) on 1 August 2015, 15h13. T $\sim$ 150s. \\
                & & & & & B & &  Faint events on 22 August 2015: 21h16, 22h16.\\
                & & & & & B & &  Faint events on 23 August 2015: 9h19, 13h11*.\\
                & & & & & B & &  Faint type B events on 30 August 2015: 6h49, 7h29, 8h09, 8h49, 9h29, 10h09, 11h41, 12h21, 13h01, 13h41, 14h21, 15h01, 16h53, 17h33, 18h13, 18h53, 19h33, 21h54, 22h34, 23h14, 23h54.\\
                & & & & & & 01h31m &  Faint events on 31 August 2015: 0h34 (A), 1h14 (B).\\
& & & & & B & &  Faint events on 31 October 2015: 10h46 (multiple components observed), 12h46, 15h07, 16h07, 17h07, 18h07, 19h07.\\
                & & & & & B & 19h52m & Bright event (B) on 31 October 2015, 20h49. T $\geq$ 12s.\\
\hline
        183 & -34.8 $\pm$ 3.5 & -30.4 $\pm$ 4.4 & Wosret & 33 & A, B, C & & Cavities B, composed of several closely placed active cavities/alcoves in the Wosret region:\\
                & & & & & A & &  Event on 26 July 2015, 12h10. T $>\sim$ 150s\\
                & & & & & & &  Faint events on 26 July 2015: 11h40 (C), 20h21 (B).\\
                & & & & & & &  Faint events on 23 August 2015: 9h19 (C), 13h11 (B), 14h11 (B).\\
                & & & & & B & & Faint events on 30 August 2015: 7h29, 8h09, 8h49, 9h29, 10h09, 11h41 ($\times$2), 13h01 ($\times$2), 14h21, 15h01, 17h33, 18h13, 18h53, 19h33 ($\times$2), 21h14, 21h54, 23h54.\\
                & & & & & B & &  Faint events on 31 August 2015: 0h34, 1h14.\\
                & & & & & A & & Faint event on 5 September 2015, 12h41*. Found on limb.\\
                & & & & & B & &  Faint events on 11 October 2015: 13h41, 16h53.\\
                & & & & & A & &  Faint event on 20 October 2015, 12h05.\\
                & & & & & B & &  Faint event on 31 October 2015, 16h07.\\
                & & & & & B & 21h27m & Bright event  on 31 October 2015, 21h49. T $>$23s\\
\hline
        184 & -40.7 $\pm$ 1.6 & -30.6 $\pm$ 2.4 & Wosret & 16 & B & & Active cavity C of the Wosret region:\\
                & & & & & & 12h14m & Bright event on 1 August 2015, 12h21.\\
                & & & & & & &  Faint events on 1 August 2015: 6h38*, 7h38, 11h21, 13h51.\\
                & & & & & & &  Faint events on 30 August 2015: 8h09, 14h21, 15h01, 18h13, 18h53, 19h33.\\
                & & & & & & &  Faint event on 5 September 2015, 16h06*.\\
                & & & & & & &  Faint event on 11 October 2015, 16h53.\\
                & & & & & & &  Faint event on 12 October 2015, 1h14*.\\
                & & & & & & &  Faint event on 20 October 2015, 5h23*.\\
                & & & & & & &  Faint event on 31 October 2015, 17h07.\\
\hline
        185 & -14.7 $\pm$ 3.1 & -25.2 $\pm$ 1.8 & Wosret & 37 & A, B, C & & Active cavity D of the Wosret region:\\
                & & & & & B & &  Bright event on 26 July 2015, 13h10. T $\sim$ 1 sequence ($\sim$ 150s).\\
                & & & & & B & & Faint events on 26 July 2015: 12h10, 13h40, 14h10.\\
                & & & & & C & & Bright  on 26 July 2015, 14h40. T $>$ 24s.\\
                & & & & & B & & Faint events on 27 July 2015, 0h13 $\&$ 1h13.\\
                & & & & & B & 00h22m & Bright event on 1 August 2015, 17h43.\\
                & & & & & & & Faint events on 1 August 2015: 6h38 (C), 7h08 (B), 8h00 (B), 8h18 (B), 13h51 (B), 15h13 (B), 15h43 (B), 16h43 (B), 17h13 (B), 18h13 (B), 18h43 (B).\\
                & & & & & A & & Clear event on 1 August 2015, 23h25. T $\sim$ 75s.\\
                & & & & & B & & Faint  events on 30 August 2015: 8h49, 9h29, 10h09, 11h41, 12h21, 13h01, 18h13, 19h33, 21h14, 21h54, 22h34, 23h14, 23h54.\\
                & & & & & B & & Faint event on 31 August 2015, 0h34\\
                & & & & & B & & Faint event on 31 October 2015, 21h49.\\
                & & & & & A & & Faint event on 20 October 2015, 11h05*.\\
                & & & & & B & & Faint event on 22 August 2015, 21h16.\\
\hline
        186 & -15.3 $\pm$ 2.8 & -36.4 $\pm$ 0.9 & Wosret & 3 & B & & Active cavity E of the Wosret region:\\
                & & & & & & 16h51m &  Faint event on 1 August 2015, 13h51.\\
                & & & & & & 02h19m &  Faint event on 31 August 2015, 0h34.\\
                & & & & & & 10h46m &  Faint event on 5 September 2015, 7h32. Found on limb.\\
\hline
        187 & -15.3 & -24.4 & Wosret & 1 & B & 23h52m &26 July 2015, 13h10. Faint.\\
 \hline
        188 & -36.9 & -33.6 & Wosret & 1 & B & 00h18m &26 July 2015, 14h10. Bright. T $>$ 23s.\\
\hline
        189 & -39.9 & -22.8 & Wosret & 1 & B & 20h35m &27 July 2015, 0h43. Faint.\\
\hline
        190 & -19.8 & -26.0 & Wosret & 1 & B & 21h01m &27 July 2015, 0h13. Faint.\\
\hline
        191 & -33.5 & -35.7 & Wosret & 1 & B & 06h29m & 1 August 2015, 9h08. Faint.\\
\hline
        192 & -22.0 & -25.2 & Wosret & 1 & B & 15h23m & 1 August 2015, 13h21. Faint.\\
\hline
        193 & 6.2 & -16.5 & Wosret & 1 & B & 00h46m & 1 August 2015, 17h43. Faint.\\
\hline
        194 & -20.1 & -21.9 & Wosret & 1 & B & 18h28m & 9 August 2015, 7h58. Faint, found on limb.\\
\hline
        195 & -39.9 $\pm$ 1.1 & -18.8 & Wosret & 2 & B, C & 17h34m-18h03m & 9 August 2015, 8h13 (B) $\&$ 8h28 (C). Faint, found on limb.\\
\hline
        196 & -37.2 & -15.1 & Wosret & 1 & A & 18h15m & 9 August 2015, 8h28. Faint, found on limb.\\
\hline
        197 & -26.1 & -12.9 & Wosret & 1 & B & 04h05m & 22 August 2015, 21h16. Faint.\\
\hline
        198 & -46.6 & -22.6 & Wosret & 1 & B & 13h39m &23 August 2015, 15h11. Clear, found on limb.\\ 
\hline
        199 & -22.4 $\pm$ 1.0 & -22.9 $\pm$ 1.7 & Wosret & 2 & B &16h25m & 30 August 2015, 7h29. Faint \\ 
           & & & & & B &01h56m &30 August 2015,12h21. Faint.\\
\hline
        200 & -9.3 & -10.6 & Wosret & 1 & B & 01h32m & 30 August 2015, 11h41. Faint.\\
\hline
        201 & -24.9 & -9.5 & Wosret & 1 & B & 03h02m & 30 August 2015, 13h01. Faint.\\
\hline
        202 & 17.1 & -16.0 & Wosret & 1 & B & 14h42m &30 August 2015, 17h33. Faint.\\
\hline
        203 & 0.0 & -18.4 & Wosret & 1 & B & 22h06m & 30 August 2015, 21h54. Faint.\\
\hline
        204 & -29.5 & -18.1 & Wosret & 1 & B & 21h25m& 30 August 2015, 22h34. Faint.\\
\hline
        205 & -40.5 & -23.4 & Wosret & 1 & B &13h20m & 31 August 2015, 0h34. Faint.\\
\hline
        206 & -44.8 & -26.2 & Wosret & 1 & B & 23h35m & 5 September 2015, 15h06. Faint.\\
\hline
        207 & -41.3 & -22.6 & Wosret & 1 & A & 21h00m & 5 September 2015, 13h41. Faint.\\
\hline
        208 & 3.9 & -15.5 & Wosret & 1 & A & 17h52m & 5 September 2015, 22h45. Faint.\\
\hline
        209 & 8.1 & -16.0 & Wosret & 1 & B & 12h15m & 5 September 2015, 7h32. Faint, found on limb.\\
\hline
        210 & -31.4 & -37.1 & Wosret & 1 & B & 00h59m & 11 October 2015, 17h53. Faint.\\
\hline
        211 & -23.5 & -26.3 & Wosret & 1 & B & 23h30m & 11 October 2015, 16h53. Faint.\\
\hline
        212 & -23.3 & -10.5 & Wosret & 1 & B & 23h50m & 11 October 2015, 17h03. Clear.\\
\hline
        213 & -20.5 & -23.2 & Wosret & 1 & B & 15h25m & 11 October 2015, 12h41. Faint.\\
\hline
        214 & -39.3 $\pm$ 1.1 & -20.7 $\pm$ 1.5 & Wosret & 1 & B & 01h06h & 31 October 2015, 23h49. Bright, relatively large event. T $\geq$ 1 sequence ($\sim$ 126s).\\
\hline
        215 & -33.3 & -18.4 & Wosret & 1 & B & 23h41m & 31 October 2015, 10h46. Faint.\\
\hline
\end{longtable}
\end{center}
}\begin{list}{}{}
\item 
\end{list}
\vspace{-2em}
\noindent

\clearpage
\twocolumn

\newpage

\end{document}